\documentclass{aa}
\pdfoutput=1  

\usepackage{graphicx}
\usepackage{txfonts}
\usepackage{xcolor}
\usepackage{placeins}
\usepackage{linenoaa}
\usepackage[colorlinks,linkcolor=blue,citecolor=blue,urlcolor=blue]{hyperref}

\begin{document} 

\title{Joining forces: 30 years of optical monitoring of the Einstein Cross}    
                                                             
\author{V. N. Shalyapin\inst{1,2,3}
                \and
                L. J. Goicoechea\inst{1,2}
                \and 
                R. Gil-Merino\inst{4}
                \and
	            A. Esteban-Guti\'errez\inst{5}
                \and
                C. W. Morgan\inst{6}
                \and
                E. Mediavilla\inst{7,8}
                \and
                A. Yonehara\inst{9} 
                \and
                A. Sergeyev\inst{10,11,12}      
                }

\institute{Instituto de F\'isica de Cantabria (CSIC-UC), Avda. de Los Castros s/n, 
		E-39005 Santander, Spain\\
                \email{vshal@ukr.net;goicol@unican.es}
                \and
		Departamento de F\'\i sica Moderna, Universidad de Cantabria, 
                Avda. de Los Castros s/n, E-39005 Santander, Spain
                \and
                O.Ya. Usikov Institute for Radiophysics and Electronics, National 
                Academy of Sciences of Ukraine, 12 Acad. Proscury St., UA-61085 
                Kharkiv, Ukraine
                \and
                Escuela Superior de Ingenier\'ia y Tecnolog\'ia, Universidad Internacional 
                de La Rioja (UNIR), Avda. Gran V\'ia Rey Juan Carlos I 41, E-26005 
                Logro\~{n}o, Spain
                \and
                Instituto de F\'isica y Astronom\'ia, Universidad de Valpara\'iso, Avda. 
                Gran Breta\~na 1111, Valpara\'iso, Chile
                \and
                Department of Physics, United States Naval Academy, 572C Holloway Rd., 
                Annapolis, MD 21402, USA
		\and
                Instituto de Astrof\'isica de Canarias, c/ V\'ia L\'actea s/n, 
                E-38205 La Laguna, Spain
		\and
                Departamento de Astrof\'isica, Universidad de La Laguna, E-38200 La Laguna, 
                Spain
                \and
                Department of Astronomy and Atmospheric Science, Faculty of Science, Kyoto 
                Sangyo University, 603-8555 Kyoto, Japan
                \and
                Universit\'e C\^ote d'Azur, Observatoire de la C\^ote d'Azur, CNRS,
                Laboratoire Lagrange, France
                \and
                Institute of Astronomy, V.N. Karazin Kharkiv National University, Sumska 35, 
                UA-61022, Kharkiv, Ukraine
                \and
                Institute of Radio Astronomy, National Academy of Science of Ukraine, 
                Mystetstv 4, Ukraine}

\abstract{We present an extended optical monitoring of the quadruply-imaged gravitationally lensed quasar QSO 2237+0305, the Einstein Cross, including observations from different observatories in both hemispheres and using a new photometric technique. This technique uses a region far enough from the lens system to determine accurately the sky background level, and minimises contamination from the lensing galaxy by combining analytical and numerical modeling of its structure. The resulting light curves of the four quasar images describe variations across practically the entire optical spectrum and span about 9000 days in the $VRI$ bands. The multi-band microlensing variability is captured with an unprecedented level of detail, and a preliminary microlensing analysis reveals an almost linear scaling of source radius with wavelength, providing direct evidence for the wavelength-dependent structure of the region contributing to optical passband fluxes. Specifically, assuming a mean microlens mass $\langle M \rangle$ = 0.3 $\rm{M_{\odot}}$ and concentric Gaussian sources that move according to the velocity distribution peaks (speed and direction) reported in a previous microlensing analysis, we find that the half-light radius of the $g$-band source is 9.6 $\pm$ 2.7 lt-day and the size of the sources grows with wavelength with a power-law index of $\alpha$ = 0.94 $\pm$ 0.05. We conclude that these long-term light curves set stringent empirical constraints on models of quasar emission and microlensing physics.}


\maketitle
\nolinenumbers

\section{Introduction}
\label{sec:intro}

The quasar \object{QSO 2237+0305} \citep[the Einstein Cross;][]{Huch85} at $z = 1.695$ is gravitationally lensed by a nearly face-on barred spiral galaxy at $z = 0.039$, producing four images (A–D) arranged in a cross configuration around the galaxy nucleus. These quasar images are seen through the bulge of the lensing galaxy, where high optical depths induce microlensing effects. In fact, soon after its discovery, the first detection of microlensing variability was made for this system \citep{Irwi89} and, from then on, reported by many other research groups \citep[e.g.,][]{Oste96,Wozn00,Eige08b,Gilm18}. 

Although it is well known that an extensive optical monitoring program of the quasar images is a critical tool to obtain strong microlensing signals, and thus robust information on the quasar accretion disc structure and galaxy bulge composition \citep[e.g.,][]{Irwi89,Wamb91,Yone98,Agol99}, such programs are very demanding and difficult to complete. Hence, several earlier efforts focused on a long-term monitoring campaign in only one bandpass filter \citep[e.g.,][]{Wozn00,Udal06} or on a relatively short multi-band campaign \citep[e.g.,][]{Eige08b,Mun16}. 

Within the framework of the Gravitational LENses and DArk MAtter (GLENDAMA) project\footnote{\url{https://gravlens.unican.es/}} \citep{Gilm18}, \object{QSO 2237+0305} has been monitored in two SDSS bandpass filters since 2006. In an earlier paper, the GLENDAMA light curves obtained with the 2.0 m Liverpool Telescope \citep[LT;][]{Ste04} in the $gr$ bands over the period 2006$-$2019 were merged with $VRI$-band brightness records from concurrent observations with the 1.5 m Maidanak Telescope (MT), illustrating the great potential of using the microlensing signal to measure the structure of the quasar accretion disc\footnote{It is commonly assumed that optical passband fluxes mostly originate in the central accretion disc} \citep{Goi20}. Here, we extend these LT-MT records by adding new $BgVrRI$ magnitudes from the LT, the MT, and the 1.3 m SMARTS Telescope (ST) over the period 1995$-$2024. Most frames have been homogeneously re-processed using a new photometric method that is described in the next Section, and the new light curves constitute the longest homogeneous multi-band dataset of \object{QSO 2237+0305} to date.

Because the inter-image time delays in \object{QSO 2237+0305} are of order hours and therefore negligible compared to typical monitoring cadences, taking the differences between image light curves (DLCs) removes the (unknown) intrinsic variations of the quasar and extracts extrinsic signals caused by microlensing \citep[e.g.,][]{Vern24}. Thus, assuming a Gaussian model to describe the multi-wavelength sources of the quasar accretion disc, multi-band DLCs spanning a few years were used to estimate the power-law index of the source radius-wavelength relationship $R_{\rm s} \propto \lambda^{\alpha}$. For example, \citet{Eige08b} reported $\alpha$ = 1.2 $\pm$ 0.3, while \citet{Mun16} found $\alpha$ = 0.7 $\pm$ 0.3. This is shallower than the standard disc model, which predicts $\alpha$ = 4/3 \citep{Shak73}, but consistent with some previous results for other lensed quasars \citep[e.g.,][]{Blac11}. 

In this work, using the new dataset spanning three decades and covering five optical bands, we discuss the inner accretion flow in \object{QSO 2237+0305} via microlensing variability. Thus, we derive constraints on the physical scale of the accretion disc and the index $\alpha$, and compare them with those from previous studies based on shorter temporal baselines. We also put our $\alpha$-index measurement into perspective regarding the current controversy over the $\alpha$ value for lensed quasars \citep[e.g.,][]{Blac11,Corn20,Sorg25}. The new dataset is a key tool not only for accretion flow analysis, but also for studying, among other things, the central engine of the quasar \citep[e.g.,][]{Best24} and the detailed composition of the lensing galaxy \citep[e.g.,][]{Tunt24,Isla25}.

The paper is organised as follows. In Sect.~\ref{sec:phot} we describe the method used to extract the fluxes of the quasar images. In Sect.~\ref{sec:lcur} we show the resulting light curves for all the facilities, cameras and bandpass filters used. In Sect.~\ref{sec:mvar} we build DLCs, measure their standard deviations, and use numerical microlensing simulations and the DLC standard deviations to constrain the size and structure of the accretion disc in \object{QSO 2237+0305}. Finally, in Sect.~\ref{sec:end} we present our conclusions.

\section{Photometric method}
\label{sec:phot}

In this Section we detail how we modelled the observed light distributions of the lens system to obtain the image light curves.    
 
\subsection{Base analytical model}
\label{sec:base}

Gravitational lenses are often modelled by a de Vaucouleurs \citep[DV;][]{Deva48} or exponential disc profile for an early or late type galaxy, respectively. Although the Einstein Cross is gravitationally lensed by a late-type spiral galaxy, its images are formed in the central bulge, which is reasonably well described by a DV profile. This approach was considered in some previous photometric schemes for analyzing small sub-frames of \object{QSO 2237+0305} \citep[e.g.,][]{Alca02,Goi20}. 

Despite the acceptable performance of a DV profile to account for the galaxy bulge, additional structures \citep[disc, central bar, and spiral arms; e.g.,][]{Spar07} also contribute to the light distribution in the central region containing the quasar images, and they must be taken into account to avoid seeing-induced variations in light contamination to the quasar images and obtain accurate photometry. For example, the galaxy disc extends beyond the central region, but has a maximum brightness in the galaxy centre. We used a base analytical model consisting of four point-like sources (quasar images) and the sum of an exponential disc and a DV bulge for the lensing galaxy, setting separations between the five sources to those measured from $Hubble$ Space Telescope imaging of the system from the CfA-Arizona Space Telescope LEns Survey (CASTLES)\footnote{CASTLES database of \object{QSO 2237+0305} at \url{https://lweb.cfa.harvard.edu/castles/Individual/Q2237.html}, updated on 21 July 2021 (C.S. Kochanek, E.E. Falco, C. Impey, J. Lehar, B. McLeod, H.-W. Rix)}. 

\subsection{Seeing effects and numerical component of the galaxy}
\label{sec:gala}

We selected large sub-frames of frames observed in good seeing conditions (typically $\sim$1\arcsec). These sub-frames are large enough to include the entire galaxy, a few bright reference stars, and a background estimation region, and we initially modelled them as a constant background plus the convolution of the base analytical model (see 
Sect.~\ref{sec:base}) with an empirical point spread function (PSF). To fit the sub-frames, we used the IMFITFITS software \citep{McLe98}, and for the empirical PSF, we considered the observed image of a bright, non-saturated star located close to the quasar. The choice of a specific PSF star depends on the optical band and seeing conditions. Sometimes the so-called PSF3 star was used and other times the PSF2 star \citep[see the finding chart in Fig. 1 of][]{Eige08a}. The two PSF stars produce similar results when both are available for photometric tasks. From the resulting fits, we determined the parameters of the analytical component of the galaxy, namely the effective radius, ellipticity, position angle, and brightness relative to a reference star for the DV and exponential disc contributions. Setting these galaxy parameters, we then computed the residual light in the selected sub-frames by subtracting solutions from our initial photometric model.

\begin{figure*}
\centering
\includegraphics[width=14cm]{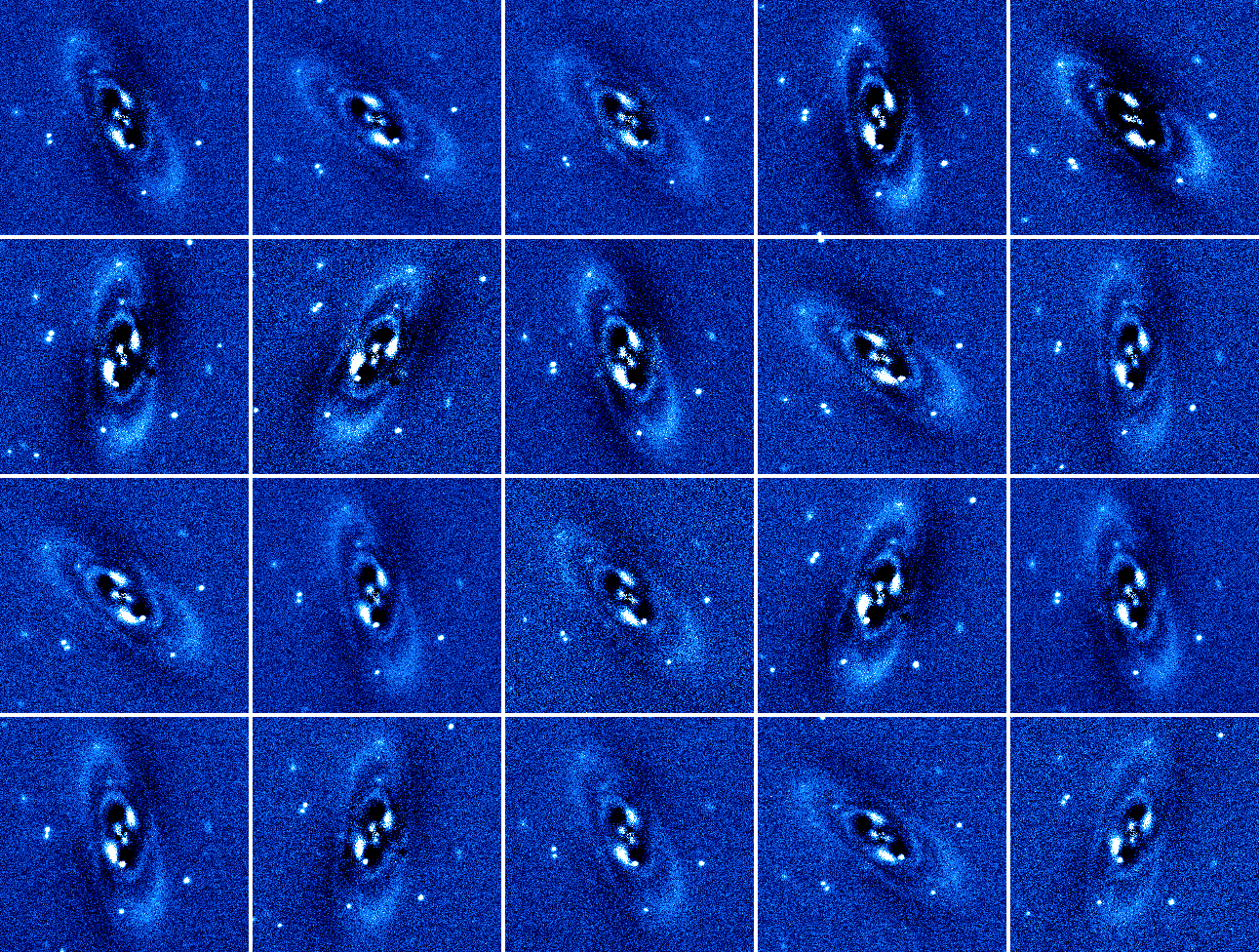}
\caption{Residual light distributions in 20 $r$-band sub-frames. These sub-frames are part of frames taken with the IO:O camera of the LT in good seeing conditions and show well-structured extra-stellar residues in the region occupied by the lensing galaxy (see main text).}
\label{fig:fig1}
\end{figure*}

\begin{figure}
\centering
\includegraphics[width=9cm]{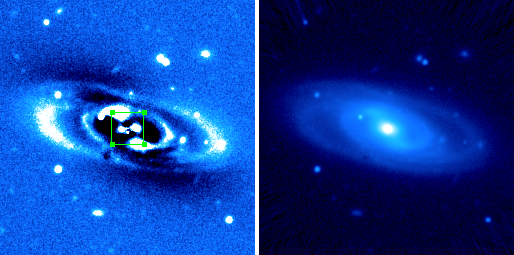}
\caption{Lensing galaxy model from the LT/IO:O/$r$-band best frames. Left panel: Numerical component in a square sub-frame of 256 pixels on a side. A 32$\times$32 pixel green square is plotted for comparison. Right panel: The sum of the analytical and numerical components.}
\label{fig:fig2}
\end{figure}

In addition to individual stars, the residual light distributions display a well-structured, non-random shape (see Figure~\ref{fig:fig1} for the best $r$-band frames in terms of seeing observed with the IO:O camera of the LT). While the frame orientation changes with time, the extra-stellar residual light shape is stable and related to structures that were not considered in the analytical component of the galaxy, i.e., the central bar, spiral arms, and possible deviations from elliptical symmetry. For each facility, camera, and filter, we combined the model residuals to create the numerical component of the galaxy as a model residuals template. The $r$-band template for the LT IO:O camera is shown in the left panel of Figure~\ref{fig:fig2}. The total galaxy model combining the analytical component and the residuals template is shown in the right panel of Figure~\ref{fig:fig2}. In comparison with previous DV-based models \citep[e.g.,][]{Alca02,Goi20}, this approach incorporates an exponential disc and the complex structures that are difficult to describe analytically. 

\begin{figure}
\centering
\includegraphics[width=9cm]{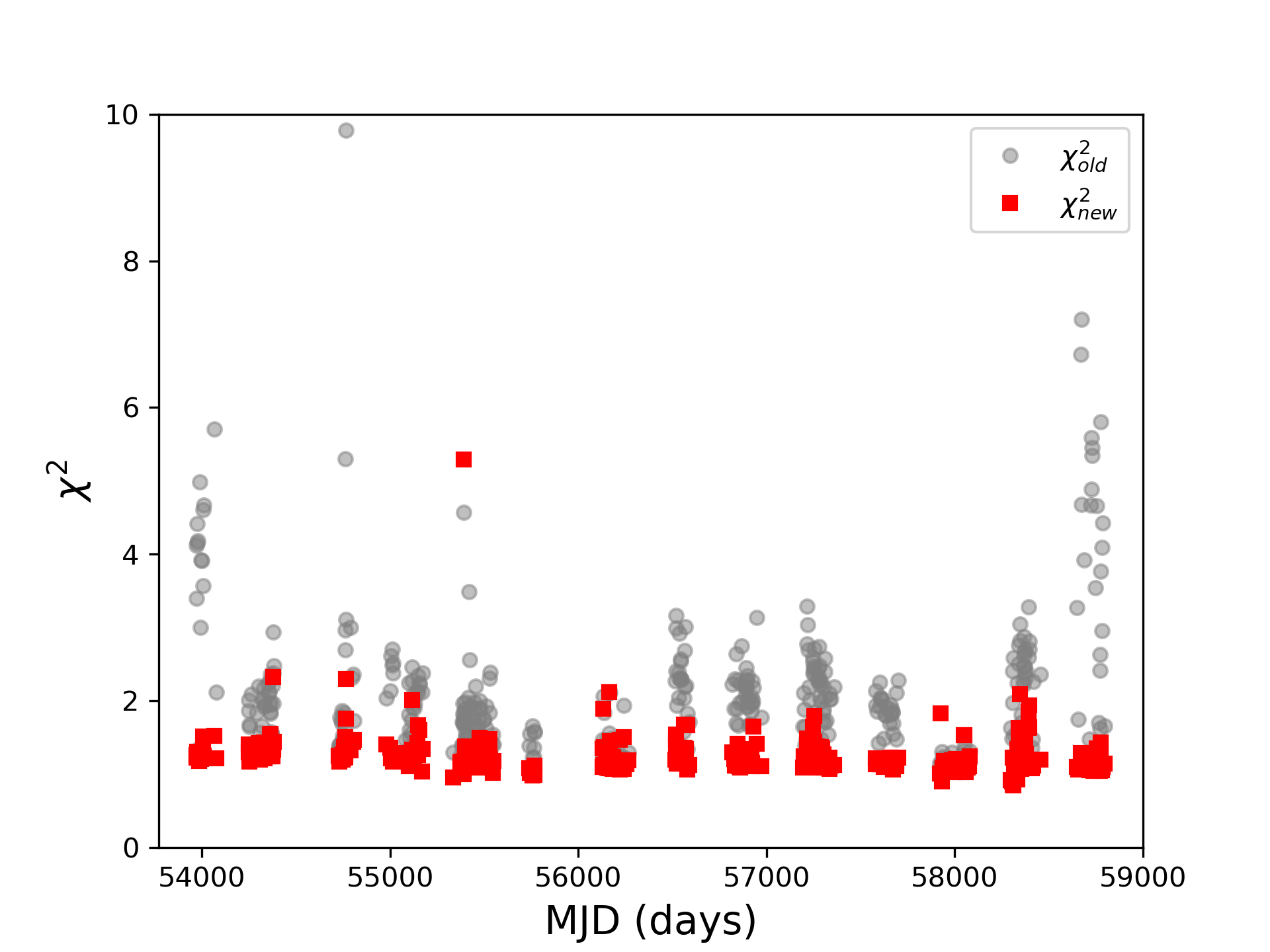}
\caption{$\chi^2$ in the central 32$\times$32 pixel square containing the four quasar images when modelling MT/$R$-band frames. The red squares correspond to the new photometric scheme and the grey circles are obtained from our previous photometric method (see main text).}
\label{fig:fig3}
\end{figure}

\subsection{Final photometric model}
\label{sec:fmod}

In general, frames were taken with seeing worse than $\sim$1\arcsec, so the numerical component of the galaxy for each observed frame was blurred to account for its PSF. We used a circular Gaussian with a $FWHM = (FWHM_{\rm obs}^2 - FWHM_{\rm best}^2)^{1/2}$ to blur the numerical part, which is good enough for our purposes because the numerical component is only a few percent of the analytical one. To obtain the quasar fluxes, the final photometric model consisted of four key ingredients: four empirical PSFs (quasar images), DV and exponential brightness distributions convolved with the empirical PSF, a numerical galaxy component convolved with the corresponding circular Gaussian function, and a constant term for the background. After estimating a calibration factor (flux of a reference star), this more detailed model has five free parameters: the sky background level and the four quasar fluxes, and it was applied to square sub-frames with 256 pixels on a side. These 256$\times$256 pixel sub-frames are much larger than those with 32 pixels on a side that were used in previous work (see the central green square in the left panel of Figure~\ref{fig:fig2}), leading to robust background estimates.

To demonstrate the good performance of the new photometric approach (using a DV + exponential disc + numerical galaxy model in 256$\times$256 pixel sub-frames) in the crowded region occupied by the quasar images, we focused on the $R$-band MT frames available for reanalysis and carried out a  goodness-of-fit test. Figure~\ref{fig:fig3} shows the new reduced chi-square ($\chi^2_{\rm new}$) in the central 32$\times$32 pixel square drawn in green in the left panel of Figure~\ref{fig:fig2}. These $\chi^2_{\rm new}$ values for 445 observing nights over 2006$-$2019 are distributed around a mean of 1.2 with a standard deviation of 0.3. For comparison, Figure~\ref{fig:fig3} also includes the $\chi^2_{\rm old}$ values from our previous photometric approach \citep[using a DV galaxy model in 32$\times$32 pixel sub-frames;][]{Goi20}, which are clearly higher and show greater dispersion (mean of 2.1 and standard deviation of 1), indicating a larger discrepancy between observed and modelled light distributions. 
 
\section{Light curves}
\label{sec:lcur}

The extended light curves of \object{QSO 2237+0305} include data obtained with three telescopes in six optical bands. The number of observing epochs (nights) for each band and telescope are shown in Table~\ref{tab:nights}. This Table~\ref{tab:nights} also includes the 
monitoring periods. It is noteworthy that the previous number of data points \citep[numbers within brackets in Table~\ref{tab:nights};][]{Goi20} is $\sim$40\% of the current one, so we increased the database by a factor of $\sim$2.4. The $B$-band exposures taken with the LT were not reprocessed with the new photometric scheme (see Sect.~\ref{sec:phot}). These exposures were so short that they yielded noisy frames that prevented an accurate estimate of the numerical galaxy model. These frames were processed with a simpler photometric method using only a DV surface brightness for the lensing galaxy. The MT frames from the period 1995$-$2005 and some MT frames in 2006$-$2008 were not available for reanalysis.

\begin{table}[ht]
\begin{center}
\caption{Number of observing epochs (nights) for each band and telescope.}
\label{tab:nights}
\begin{tabular}{lccccc}
\hline\hline
Band    & \multicolumn{4}{c}{Epochs/nights} &   Period           \\
\cline{2-5}                                                  
        &  LT       &    MT    &    ST      & Total &            \\
\hline                                                       
$B$     &  70       &    11    &   152      &  233  & 2004$-$2020\\
$g$     & 319(203)  &          &            &  319  & 2006$-$2024\\
$V$     &  66       & 335(180) &    43      &  444  & 1995$-$2020\\
$r$     & 372(253)  &          &            &  372  & 2006$-$2024\\
$R$     &           & 843(445) &   336      & 1179  & 1995$-$2019\\
$I$     &           & 308(179) &   146      &  454  & 1995$-$2019\\
\hline 
\end{tabular}
\end{center}
\footnotesize{Note: The last two columns show the total number of epochs and the monitoring period for each band. Numbers in brackets refer to the number of epochs analysed in \citet{Goi20}.}
\end{table} 

\subsection{LT monitoring in the $BgVr$ bands}

\begin{figure*}
\centering
\includegraphics[width=14cm]{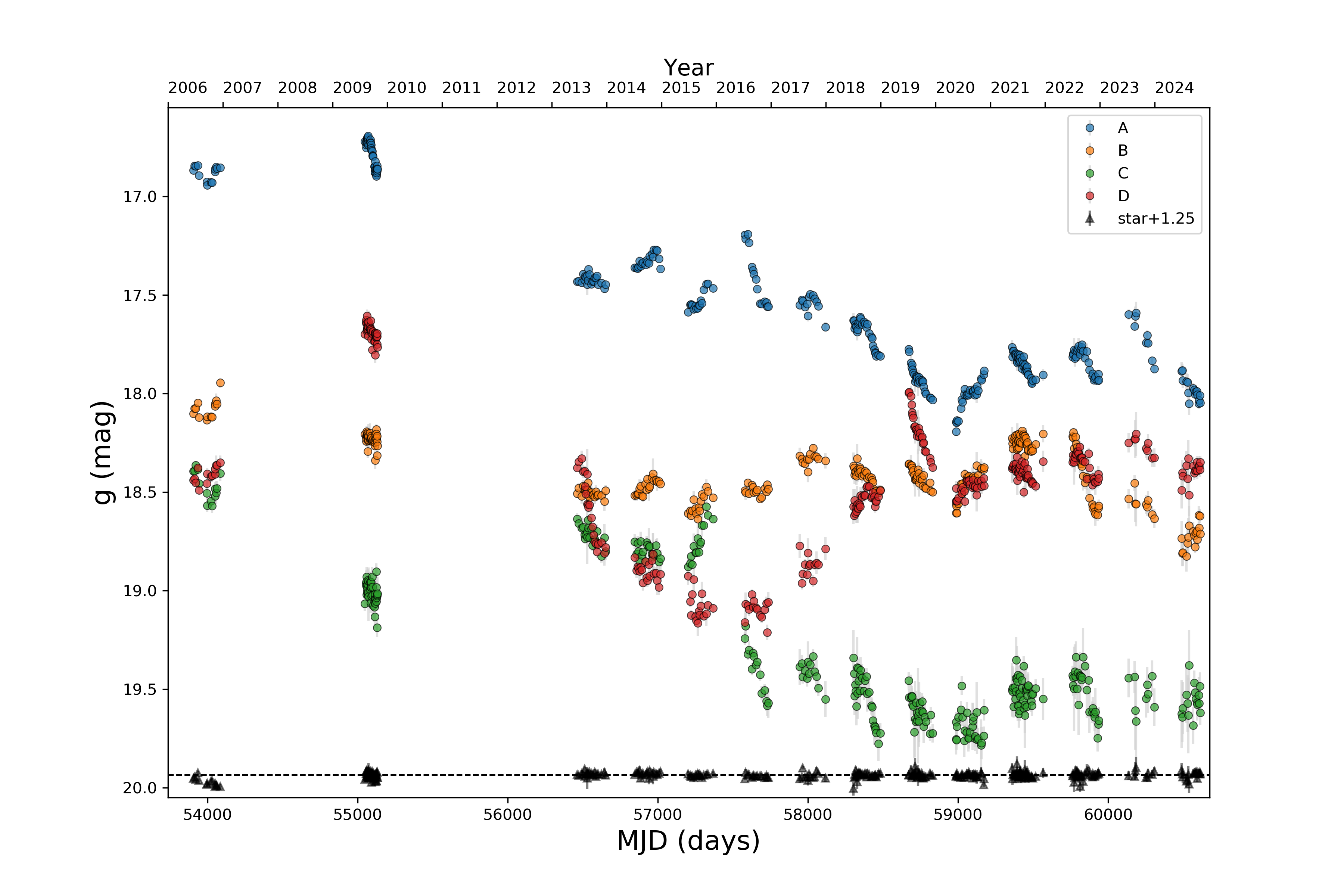}
\includegraphics[width=14cm]{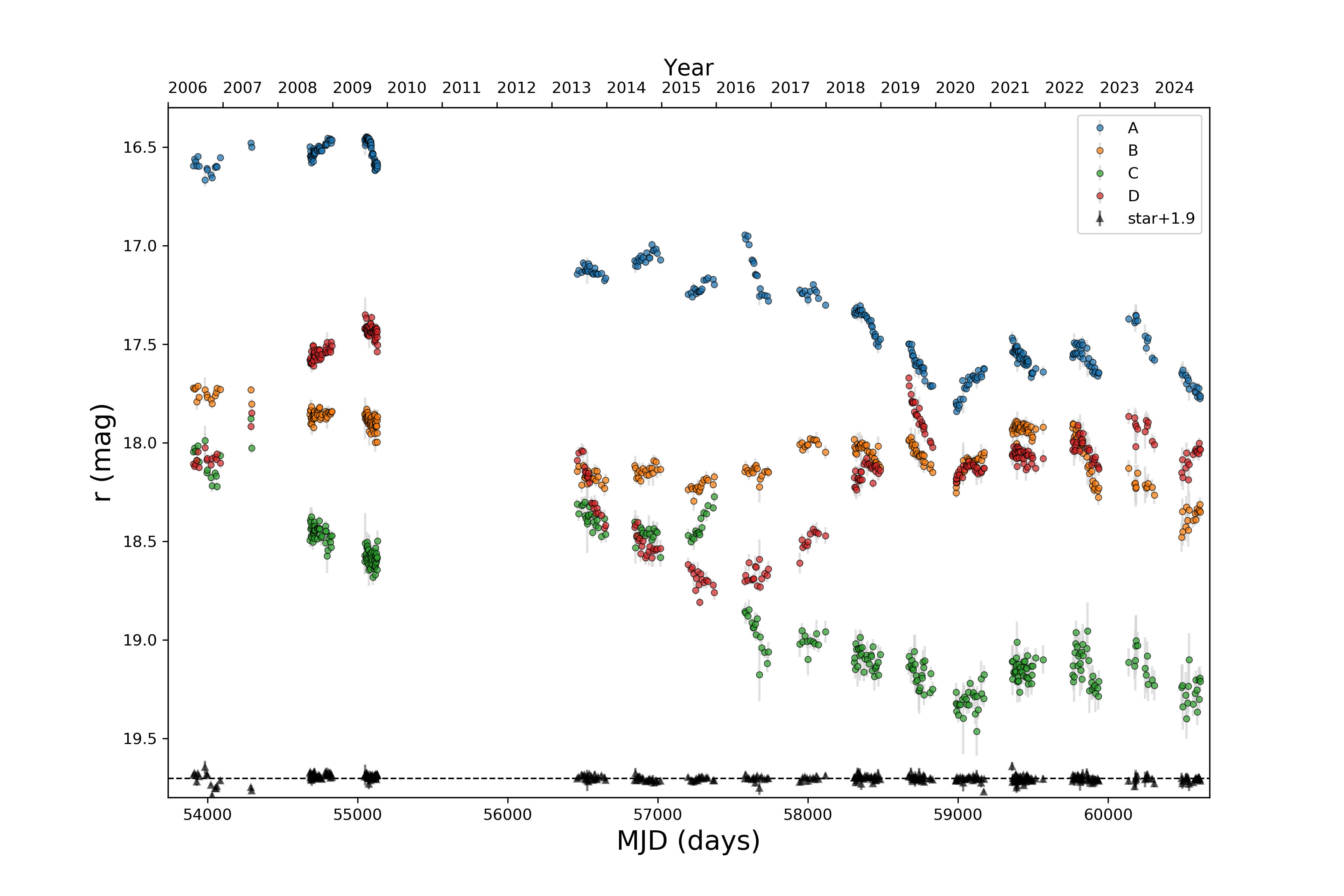}
\caption{Updated GLENDAMA light curves of \object{QSO 2237+0305}. We added new data resulting from monitoring with the LT over 2020$-$2024. Top panel: $g$-band magnitudes. Bottom panel: $r$-band magnitudes.}
\label{fig:fig4}
\end{figure*}

Monitoring in the $gr$ bands was carried out with two different CCD cameras: RATCam (0\farcs27/pixel) over 2006$-$2009 and IO:O (0\farcs30/pixel) over 2013$-$2024. Each IO:O observing night, we took one 300 s (180 s) exposure in the $g$ ($r$) band. The updated GLENDAMA light curves for the $g$ and $r$ bands are presented in the top and bottom panels of Figure~\ref{fig:fig4}, respectively, and in Tables C.1 ($g$ band) and C.2 ($r$ band). These long photometric tables are available at the CDS in a standard format containing epochs, magnitudes, and magnitude errors. The new photometric method mainly produces small increases in brightness in the light curves of the C and D images compared to our earlier light curves, with larger increases when the image is fainter. In addition to these magnitude offsets probably related to adequate subtraction of the sky background, the scatter is reduced, presumably due to our improved modelling of the observed light distributions. The top and bottom panels of Figure~\ref{fig:fig4} also show that the four images roughly present similar variations in the last 5 years. We also note the presence of a significant extrinsic variation in image D in 2019.

\begin{figure*}
\centering
\includegraphics[width=\textwidth]{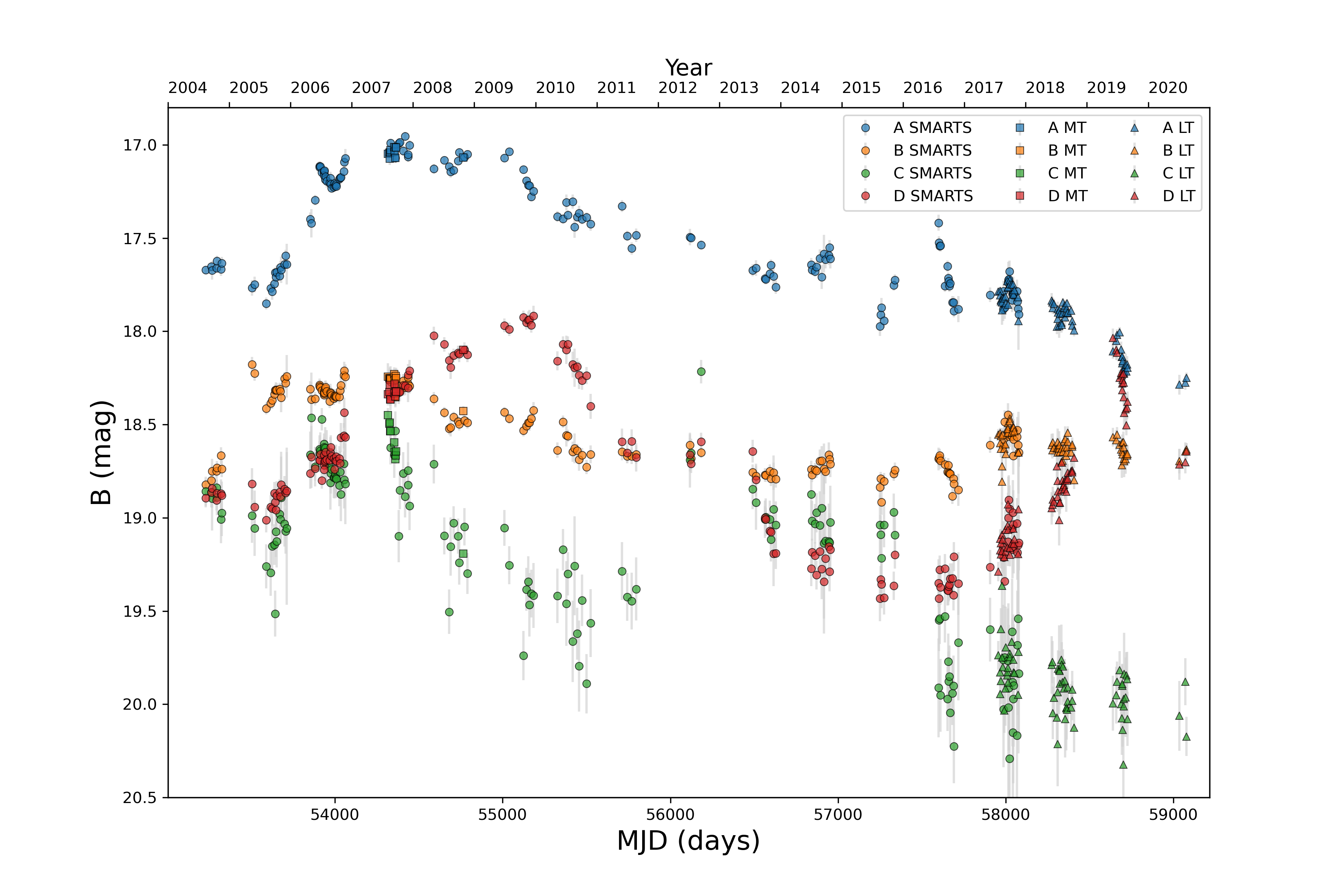}
\includegraphics[width=\textwidth]{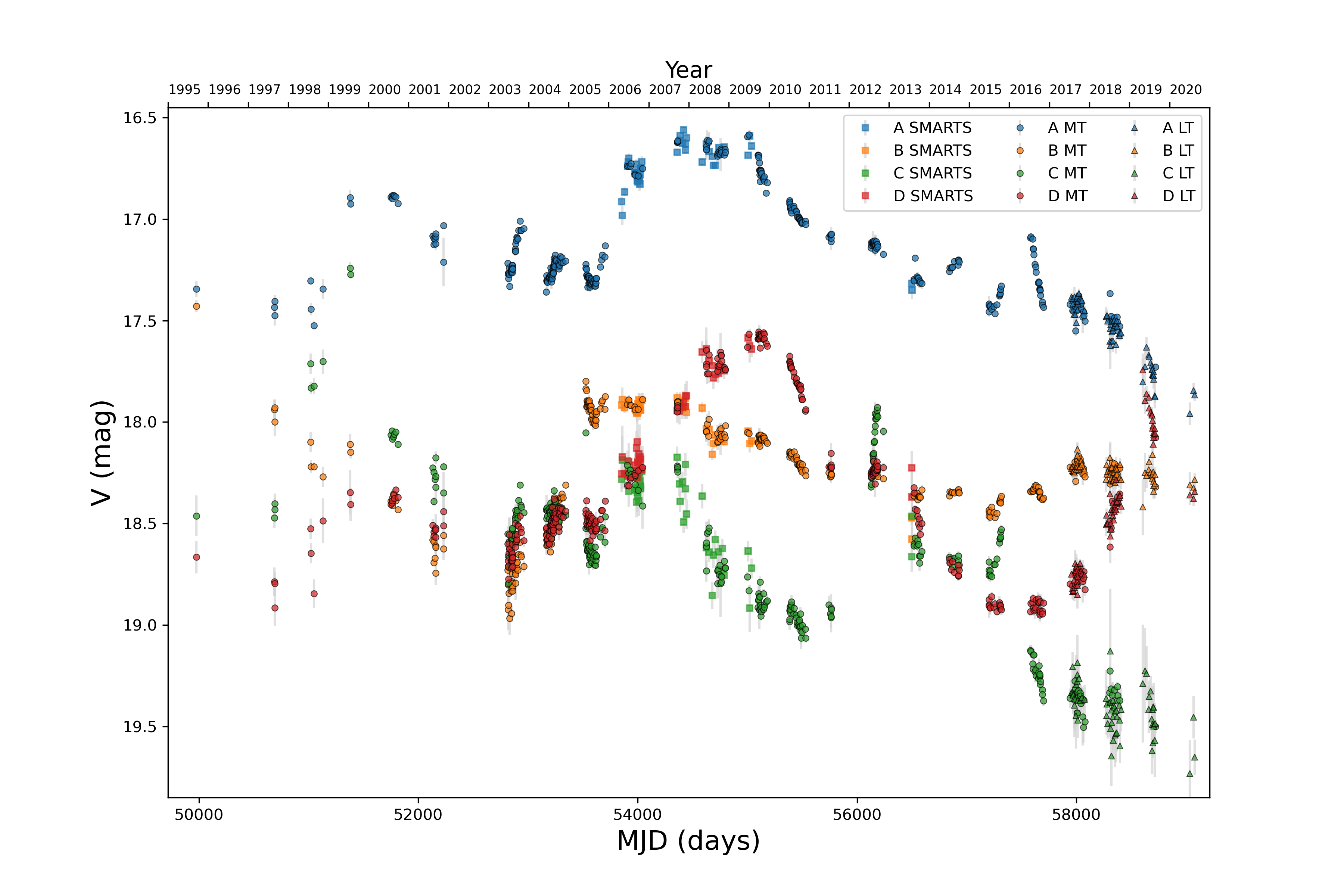}
\caption{Updated $B$-band ({\it top panel}) and $V$-band ({\it bottom panel}) light curves.}
\label{fig:fig5}
\end{figure*}

\begin{figure*}
\centering
\includegraphics[width=\textwidth]{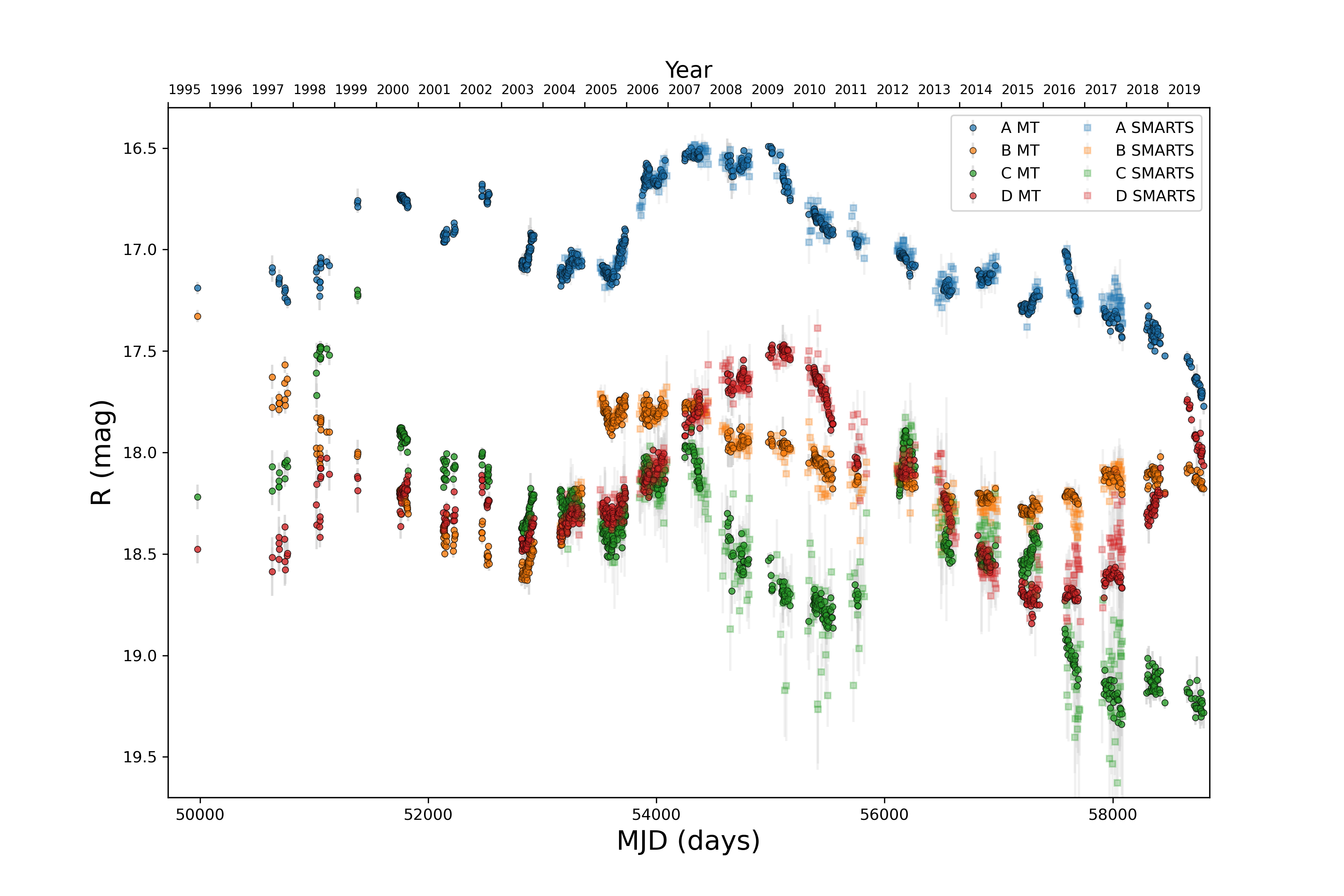}
\includegraphics[width=\textwidth]{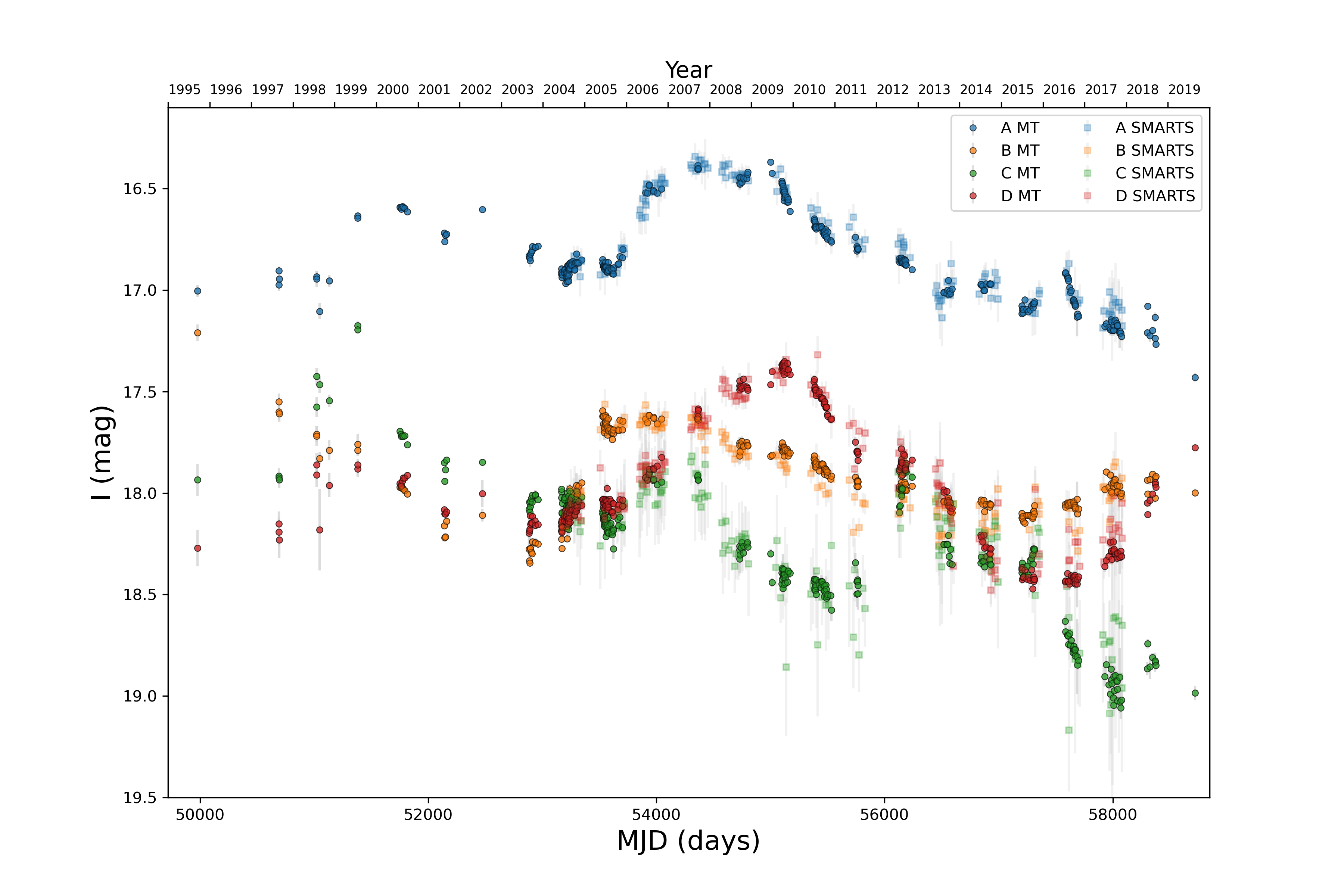}
\caption{Updated $R$-band ({\it top panel}) and $I$-band ({\it bottom panel}) light curves.}
\label{fig:fig6}
\end{figure*}

\begin{figure*}
\centering
\includegraphics[width=\textwidth]{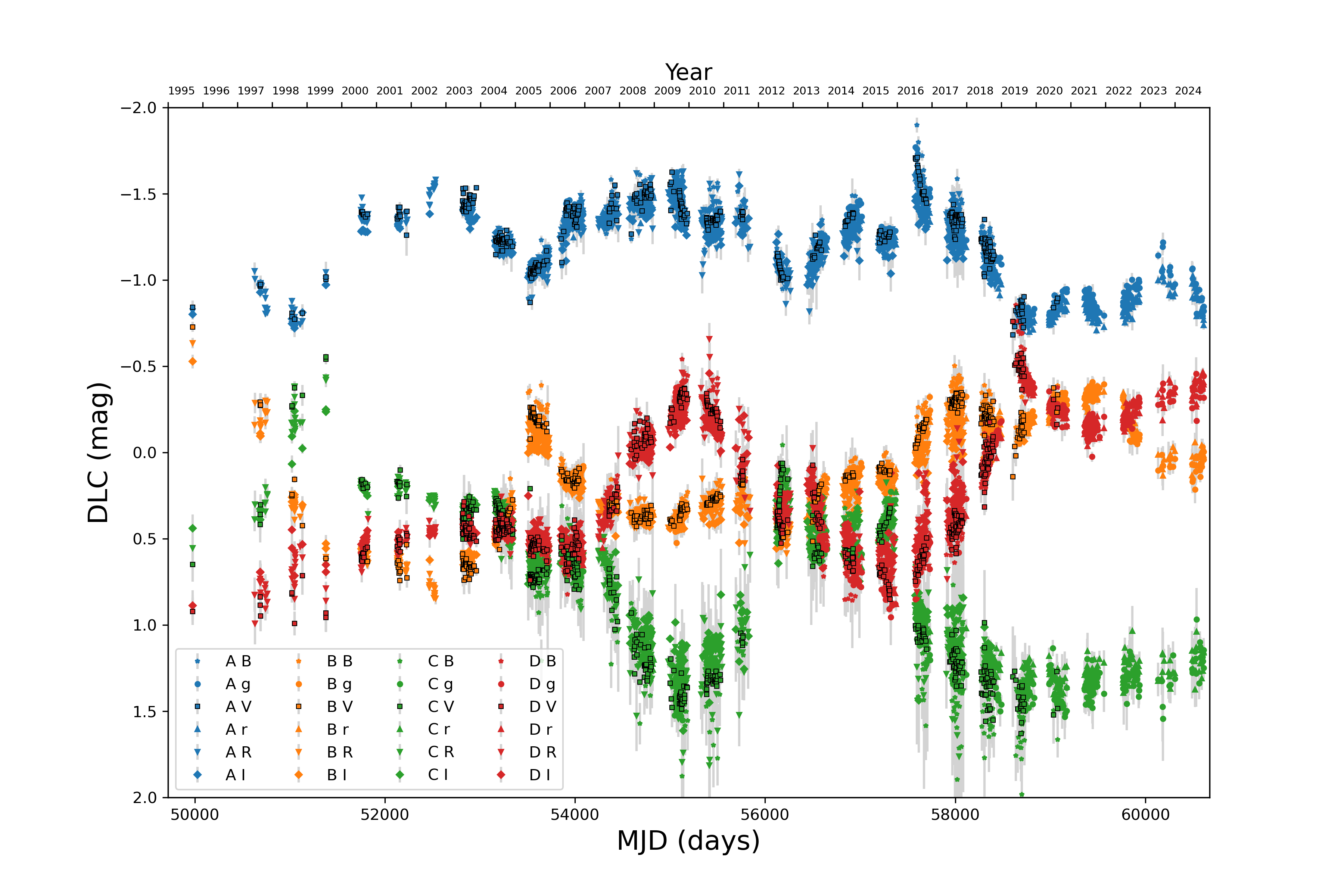}
\caption{Difference light curves in the $BgVrRI$ bands. Squares with black borders draw the microlensing signals in the $V$ band (see main text). Details of the DLCs in five well-sampled observing seasons are shown in Appendix~\ref{sec:zoom2}.}
\label{fig:fig7}
\end{figure*}

We also analysed the LT/IO:O frames in the $BV$ bands. This program typically consisted of 4$\times$20 s $B$-band exposures and 4$\times$25 s $V$-band exposures every observing night. The signal-to-noise ratio of the short exposures in the $B$ band is lower than those for the 
$gr$ bands, but allowed us to extend the $B$-band monitoring period by three additional years.

\subsection{MT monitoring in the $BVRI$ bands}

MT frames in 2006$-$2019 were obtained with two CCD cameras: SNUCAM (0\farcs27/pixel) and FLI MicroLine (0\farcs21/pixel). The average exposure time per night was relatively long (1290 s) and the average seeing per night was reasonably good (1\farcs25). The $R$-band light curve is the best sampled, with lower sampling rates for the $VI$ bands and very poor sampling for the $B$ band (see Figures~\ref{fig:fig5} and \ref{fig:fig6}). After reprocessing all available MT frames, we realized that the new refined photometry for 2006$-$2008 follows closely previously published MT brightness records for that period \citep{Dudi10}. We only found small constant magnitude offsets between the two. Hence, although the magnitudes of \citet{Dudi10} from 2001 to 2008 are based on a photometric method different from ours, we used their light curves for the period 2001$-$2008 after correcting for the magnitude offsets, and then extended the MT records by six more years (1995$-$2000) using magnitudes from the same group obtained through photometric approaches consistent with our refined photometry \citep{Vaku04}. This dataset includes only one night in 1995~\textemdash\ regular monitoring only started in 1997.

\subsection{ST monitoring in the $BVRI$ bands}

The ST $BVRI$ observations were performed with the ANDICAM camera 
\citep[0\farcs37/pixel;][]{DePoy03} over the period 2004$-$2017. Typically, three 300 s exposures per night were taken in each filter, and basic instrumental reductions were applied to all ST frames (i.e., bias subtraction, trimming of the overscan regions, flat-fielding, and cosmic-ray cleaning). The frames were processed with our new photometric method to extract the quasar magnitudes. To obtain the final combined $VRI$ light curves (see the three corresponding panels of Figures~\ref{fig:fig5} and \ref{fig:fig6}), we considered the MT magnitudes as a reference, and corrected the LT and ST data for small constant magnitude offsets. However, the number of MT/$B$-band epochs is small (see Table~\ref{tab:nights}), so we used the ST $B$-band magnitudes as a reference to build the combined light curves in that optical band, correcting the LT and MT data with small magnitude offsets (see the top panel of Figure~\ref{fig:fig5}). These data are given in Tables C.3 ($B$ band), C.4 ($V$ band), C.5 ($R$ band), and C.6 ($I$ band) at the CDS, with a last column indicating the facility. 

\subsection{Error estimation and short-term variability}

For each band and telescope, we estimated typical magnitude errors of the four quasar images and a control star. The PSF1 star in Fig. 1 of \citet{Eige08a} \citep[it is the so-called $\beta$ star in Fig. 1 of][]{Corr91} was used as a control object having a constant true magnitude in each optical band. For a given band, the true magnitude of PSF1 is close to the average magnitude of some quasar image, so the PSF1 star is well-suited for checking the reliability of typical magnitude errors of the quasar in that band. For example, if we focus on the MT $R$-band data, the true magnitude of PSF1 ($\sim$18 mag) is similar to the average $R$-band magnitude of B. Thus, for each quasar image and the PSF1 star, we calculated the standard deviation between MT $R$-band magnitudes with time separations $\leq$2.5 d, and then divided it by the square root of 2 to obtain the typical error. As expected, the typical errors of image B and the control star have similar values and amount to $\sim$2\%, while we achieve $\sim$3\% photometry from MT $R$-band observations of the image C, showing episodes with a brightness $\sim$1 mag greater than that of the control star.

For each quasar image, uncertainties at every epoch were also estimated by weighting its typical error by the $\langle S/N \rangle / S/N$ ratio \citep[e.g.,][]{Howe06}, where $S/N$ is the signal-to-noise ratio and $\langle S/N \rangle$ is the average $S/N$. It is worth noting that these individual photometric uncertainties are distributed around typical errors, which significantly exceed theoretical ones, based on expected noise sources in image formation and excluding dominant systematic issues.

The short-term multi-band variability over three well-sampled time periods spanning one or two years is shown in Figures~\ref{fig:lcs2013}, \ref{fig:lcs2015_16}, and \ref{fig:lcs2018_19} in Appendix~\ref{sec:zoom}. These zoomed-in multi-band light curves clearly display intra-year and year-to-year chromatic variations exceeding the observational noise. For example, in 2013 (see Figure~\ref{fig:lcs2013}), the optically quiescent image B is compared with the optically active image D. On a timescale of $\sim$100 d, the brightness of D decreases by $\sim$0.5 mag in the $Bg$ blue bands, but only by $\sim$0.3 mag in the $R$ band. Additionally, in 2015$-$2016 (see Figure~\ref{fig:lcs2015_16}), the brightness of C drops by $\sim$1 mag in $\sim$1 yr in the $g$ band. However, the brightness gradient in C is appreciably less pronounced in the $I$ band. 

\section{Microlensing variability}
\label{sec:mvar}

\subsection{Difference light curves}
\label{sec:dlcs}

Since the time delays between the quasar images in \object{QSO 2237+0305} are less than one day \citep[e.g.,][]{Dai03,Vaku06,Berd17}, the observed microlensing signals are simply the differences between light curves, usually using image pairs \citep[e.g.,][]{Vern24}. Here, we used four non-standard DLCs in each optical band: A $-$ (B + C + D)/3, B $-$ (A + C + D)/3, C $-$ (A + B + D)/3, and D $-$ (A + B + C)/3, which are called DLC A-D and are shown in Figure~\ref{fig:fig7} (see also Appendix~\ref{sec:zoom2}). These DLCs fully remove the intrinsic variability of the quasar and minimise the microlensing contribution from the three subtracted curves to the signal in the first.

In the absence of microlensing effects produced by stars, the DLCs would be constant but band dependent due to extinction differences. The DLCs in Figure~\ref{fig:fig7} clearly show strong microlensing-induced variability that also depends on the band, since microlensing signal amplitudes are related to the source size which depends on the wavelength \citep[e.g.,][]{Schn92,Vern24}. Isolated high-magnification events are seen, for example, in the DLC A in 2016, DLC B in 2005, DLC C in 1999, and DLC D in 2019. Moreover, the DLC C shows evidence of a double caustic-crossing event covering the period 2011$-$2016, first detected by \citet{Goi20} and currently being analysed through detailed numerical simulations. 

\begin{figure}
\centering
\includegraphics[width=9cm]{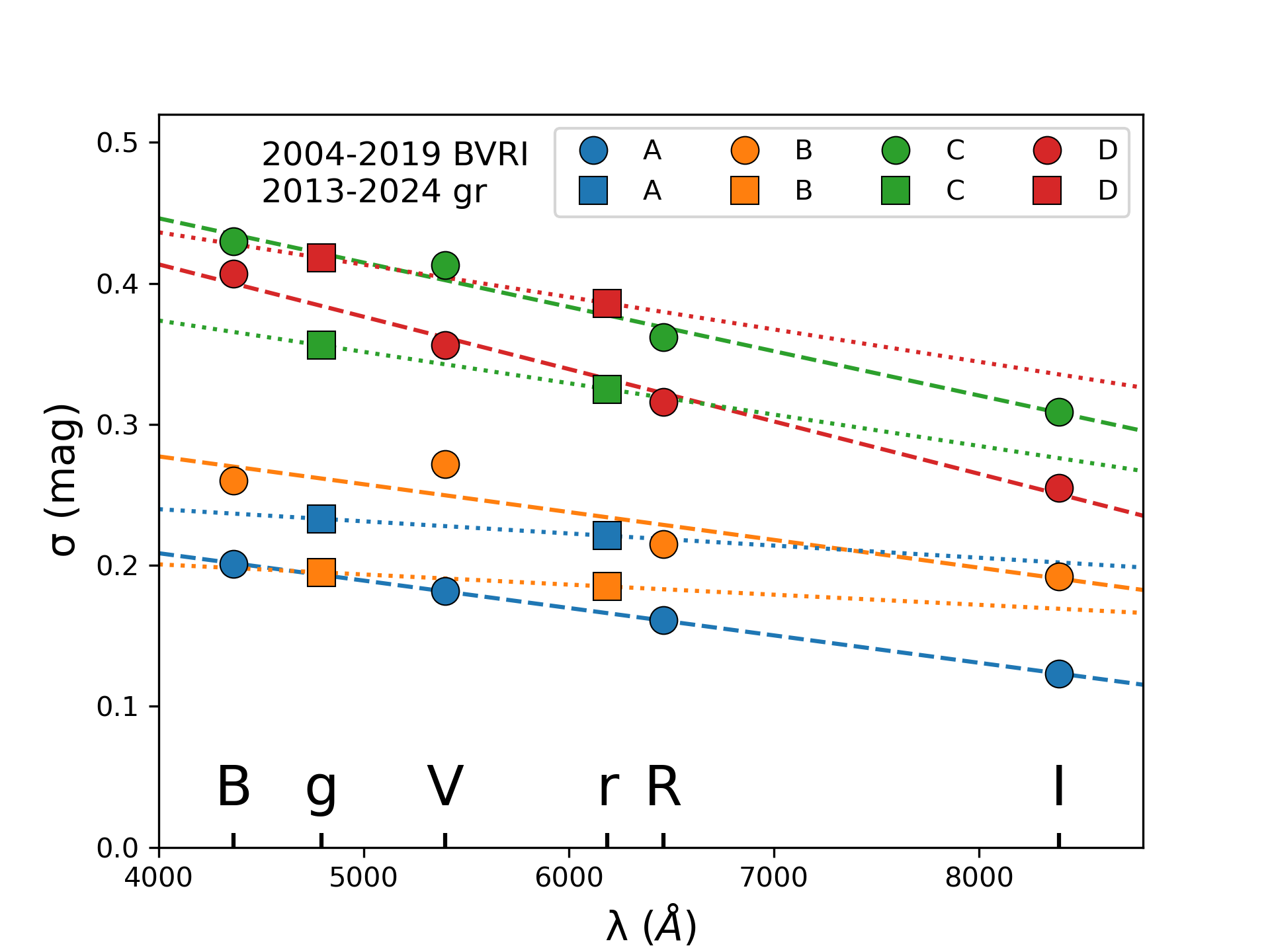}
\caption{Root mean square microlensing variability for all images as a function of wavelength. We also show a linear fit for each quasar image and gap-free time period (dashed lines).}
\label{fig:fig8}
\end{figure}

As a first approach to quantifying microlensing, we computed the root mean square dispersion $\sigma$ for each image and band over well-sampled time periods (2004$-$2019 for $BVRI$ and 2013$-$2024 for $gr$; see Figures~\ref{fig:fig4}, \ref{fig:fig5}, and \ref{fig:fig6}). The values are in Table~\ref{tab:sigm} and shown in Figure~\ref{fig:fig8}. Standard error propagation yielded uncertainties of a few percent in the $\sigma$ values, which are not included in Table~\ref{tab:sigm}, but are considered in Sect.~\ref{sec:stru}. In Figure~\ref{fig:fig8} we see that images C and D have significantly higher microlensing variability than images A and B, consistent with theoretical predictions based on the convergence, shear, and compact-object mass fraction (stars) at the image positions \citep[e.g.,][]{Seit94,Lewi96}. Additionally, microlensing variability is lower at longer wavelengths as expected for a source size-wavelength relationship $R_{\rm s} \propto \lambda^{\alpha}$ with $\alpha >$ 0. From X-ray observations of a sample of quadruply-imaged quasars, \citet{Guer20} showed that a comparison between the measured values of $\sigma$ and those of simulated DLCs may provide relevant information on the X-ray sources. Thus, we explore here a similar approach from our optical DLCs. 

\begin{table}
\centering
\caption{Standard deviations of the DLCs in mag.}
\begin{tabular}{lccccc}
\hline\hline
Band    &    A    &    B    &    C   &    D    &     Period   \\
\hline                                                       
$B$     &  0.201  &  0.260  &  0.430 &  0.407  &   2004$-$2019\\
$g$     &  0.233  &  0.197  &  0.358 &  0.418  &   2013$-$2024\\
$V$     &  0.182  &  0.272  &  0.413 &  0.356  &   2004$-$2019\\
$r$     &  0.221  &  0.186  &  0.328 &  0.385  &   2013$-$2024\\
$R$     &  0.161  &  0.215  &  0.362 &  0.316  &   2004$-$2019\\
$I$     &  0.123  &  0.192  &  0.309 &  0.255  &   2004$-$2019\\
\hline 
\end{tabular}
\label{tab:sigm}
\end{table}

For a more quantitative analysis, we created magnification maps for the quasar images, using the convergences, shear strengths, and microlens mass fractions from Table 1 of \citet{Guer20}. The stars were randomly selected from a mass function $dN/dM \propto M^{-1.3}$ with a maximum-to-minimum mass ratio of 50 \citep{Goul00}. We generated source-plane magnification maps for a point-like source (one pixel size) using the Poisson and Inverse Polygon (PIP) method\footnote{https://microlensing.overfitting.es/} 
\citep{Shal21}. This publicly available software makes map generation quick and simple. The maps were convolved with Gaussian kernels $\exp(-r^2/2R^2)$, where $R_{\rm s}$ = 1.177 $R$ is the half-light radius of the source in units of the Einstein radius of a star with a mean mass $\langle M \rangle$ ($R_{\rm E}$). 

To account for random motions of stars, in this paper, we created dynamic maps \citep{Wamb95,Poin10a,Poin10b} that we call magnification {\it cubes}. These animated sequences of static magnification patterns were created assuming a typical ratio of 2.5 between the effective speed of the source in the lens plane $v_{\rm t}$ \citep[its probability distribution has a peak around 400 km s$^{-1}$; see Fig. 5 of][]{Poin10a} and the stellar 
velocity dispersion of the lensing galaxy $\sigma_* \sim$ 160$-$170 km s$^{-1}$ \citep{Trott10,vandV10}. Rather than using a Gaussian distribution of stellar velocities, we simply treated the stellar motions in the lens plane as a distribution of velocities with the same modulus $\sigma_*$ and arbitrary directions. 

\begin{figure}
\centering
\includegraphics[width=9cm]{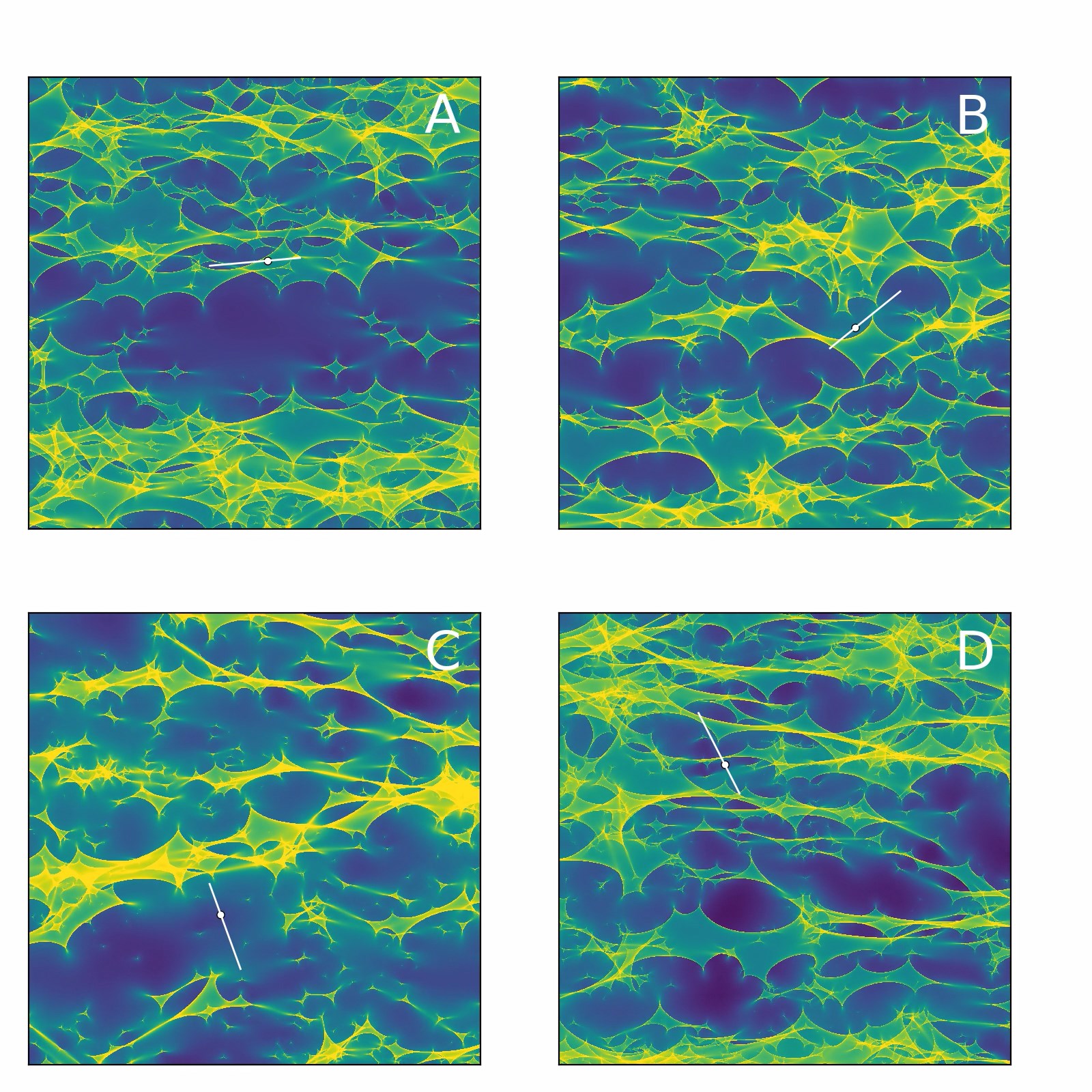}
\caption{{\bf (Movie online)} Animated sequence of magnification maps corresponding to 25 simulation steps of 0.2 $R_{\rm E}$ each. The white line is the source trajectory across the sky, which is rotated in the B, C, and D magnification {\it cubes} with respect to {\it cube} A because of misalignments between the coordinate systems. For each quasar image, the shear direction coincides with the X-axis on the corresponding magnification map, and the four shear directions form different angles with celestial north.}
\label{fig:fig9}
\end{figure}

We considered magnification maps 25 $R_{\rm E}$ on a side and a resolution of 2500$\times$2500 pixels (i.e., 100 pix/$R_{\rm E}$), and checked that larger maps produce similar results. For each optical band, we produced simulated light curves for the four quasar images with the same time sampling as the observed ones. We took random starting points on each first image map (first map of the image {\it cube}) and used the same source trajectory for all images. Each trajectory has a length $L$ and a direction depending on the shear orientation of the image \citep{Witt94,Poin10a}. Figure~\ref{fig:fig9} displays an example with $L$ = 5 $R_{\rm E}$ for the four images. We used the PIP software and a step of 0.2 $R_{\rm E}$ for the source displacement, which means that the stars were displaced by 0.08 $R_{\rm E}$ at each simulation step. This source-to-star displacement ratio of 2.5 coincides with the $v_{\rm t}/\sigma_*$ ratio.

\begin{figure*}
\sidecaption
\includegraphics[width=12cm]{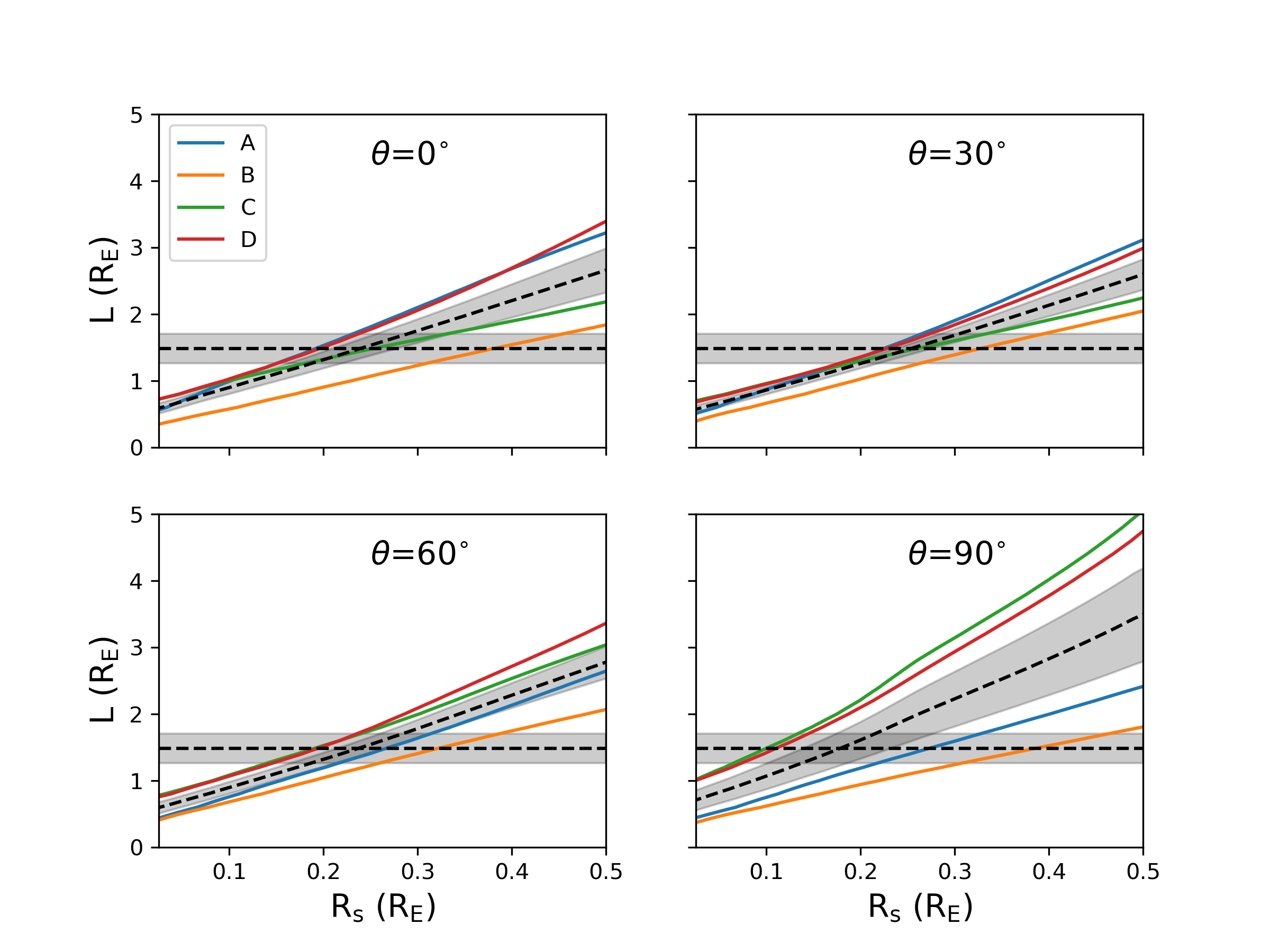}
\caption{$R_{\rm s}$-$L$ curves consistent with the standard deviations of the observed $g$-band DLCs. We consider four trajectory angles measured east of north ($\theta$ = 0, 30, 60, and 90\degr). The inclined black dashed lines and the gray regions around them describe the average $R_{\rm s}$-$L$ curves and their standard error bands, while the horizontal black dashed line and the horizontal gray strip are associated with an estimate of the trajectory length (see main text).}
\label{fig:fig10}
\end{figure*} 

For each source trajectory, we obtain four simulated light curves that allow us to build synthetic DLCs. There are constant offsets between the observed and synthetic DLCs because our simulations ignore dust extinction effects, but these do not affect the DLC dispersions. The amount of microlensing variability of a quasar image (standard deviation of its DLC) is related to the source radius $R_{\rm s}$ (the larger the source, the smaller the variations) and the trajectory length $L$ (longer trajectories have larger variations). The source trajectory angle $\theta$ (measured east of north) also plays a role \citep[e.g.,][]{Wamb90}. For each optical band and quasar image, we searched for ($R_{\rm s}$, $L$, $\theta$) values leading to a simulated standard deviation $\sigma_{\rm sim}$ (average for a representative set of starting points) equal to the observed one $\sigma_{\rm obs}$. 

\subsection{Accretion disc structure}
\label{sec:stru}

In Figure~\ref{fig:fig10}, we focus on the $g$-band data (see the $\sigma_{\rm obs}$ values in the third row of Table~\ref{tab:sigm}) and try to estimate the source size. For the Einstein Cross, several previous studies indicated that $R_{\rm s}$ at optical wavelengths is appreciably smaller than $R_{\rm E}$ \citep[e.g.,][]{Koch04,Angu08,Poin10a,Poin10b}. 
Hence, we considered a uniform distribution of 100 values of $R_{\rm s}$, ranging from 0.01 $R_{\rm E}$ to $R_{\rm E}$. We also tested 25 values of $L$, distributed uniformly between 0.2 and 5 $R_{\rm E}$. We used four source trajectory angles ($\theta$ = 0, 30, 60, and 90\degr). For every possible combination of $R_{\rm s}$, $L$, and $\theta$, we generated 1000 source trajectories, built the corresponding synthetic DLCs, calculated their standard deviations, and derived $\sigma_{\rm sim}$ for each image by averaging the 1000 individual results. Thus, for each image, setting $\sigma_{\rm sim} = \sigma_{\rm obs}$ at a given value of $\theta$, we obtain a set of pairs ($R_{\rm s}$, $L$) consistent with the observations. The 16 resulting $R_{\rm s}$-$L$ curves (4 images $\times$ 4 source trajectory directions) are shown in Figure~\ref{fig:fig10}.

With respect to the direction of motion of the source, the four $R_{\rm s}$-$L$ curves for an angle of 30\degr\ east of north are slightly closer to each other than those for the 60\degr\ angle, but clearly closer together than for the other two values of $\theta$. Additionally, in Fig. 4 of \citet{Poin10a}, $\theta$ has its probability peak in the interval 30$-$60\degr. By selecting the four coloured lines in the top right panel of Figure~\ref{fig:fig10} ($\theta$ = 30\degr), we can construct an average curve and calculate its standard error band (black dashed line and gray region around it). We can also estimate the trajectory length of the $g$-band source. We adopted a concordance flat cosmology and a standard mean mass $\langle M \rangle$ = 0.3 $\rm{M_{\odot}}$, yielding $R_{\rm E}$ = 10$^{17}$ cm \citep[e.g.,][]{Poin10a,Poin10b}. As the $g$-band data span about 11.5 yr, $L \sim$ 1.5 $R_{\rm E}$ for a typical $v_{\rm t}$ of $\sim$400 km s$^{-1}$ (see above). We assumed that the actual $v_{\rm t}$ is within a 15\% interval around its typical value, that is, the true transverse speed can be up to 15\% lower or higher than the typical one (gray region around the horizontal black dashed line). This speed interval (400 $\pm$ 60 km s$^{-1}$) agrees with the speed distribution peak in Fig. 5 of \citet{Poin10a}. The intersection between the two gray regions in the top right panel of Figure~\ref{fig:fig10} yields $R_{\rm s}$ = (0.25 $\pm$ 0.07) $R_{\rm E}$ for the half-light radius of the $g$-band source. An analysis of the results in the bottom left panel of Figure~\ref{fig:fig10} ($\theta$ = 60\degr) led to a measurement of $R_{\rm s}$ in good agreement with that for $\theta$ = 30\degr.

\begin{figure*}
\sidecaption
\includegraphics[width=12cm]{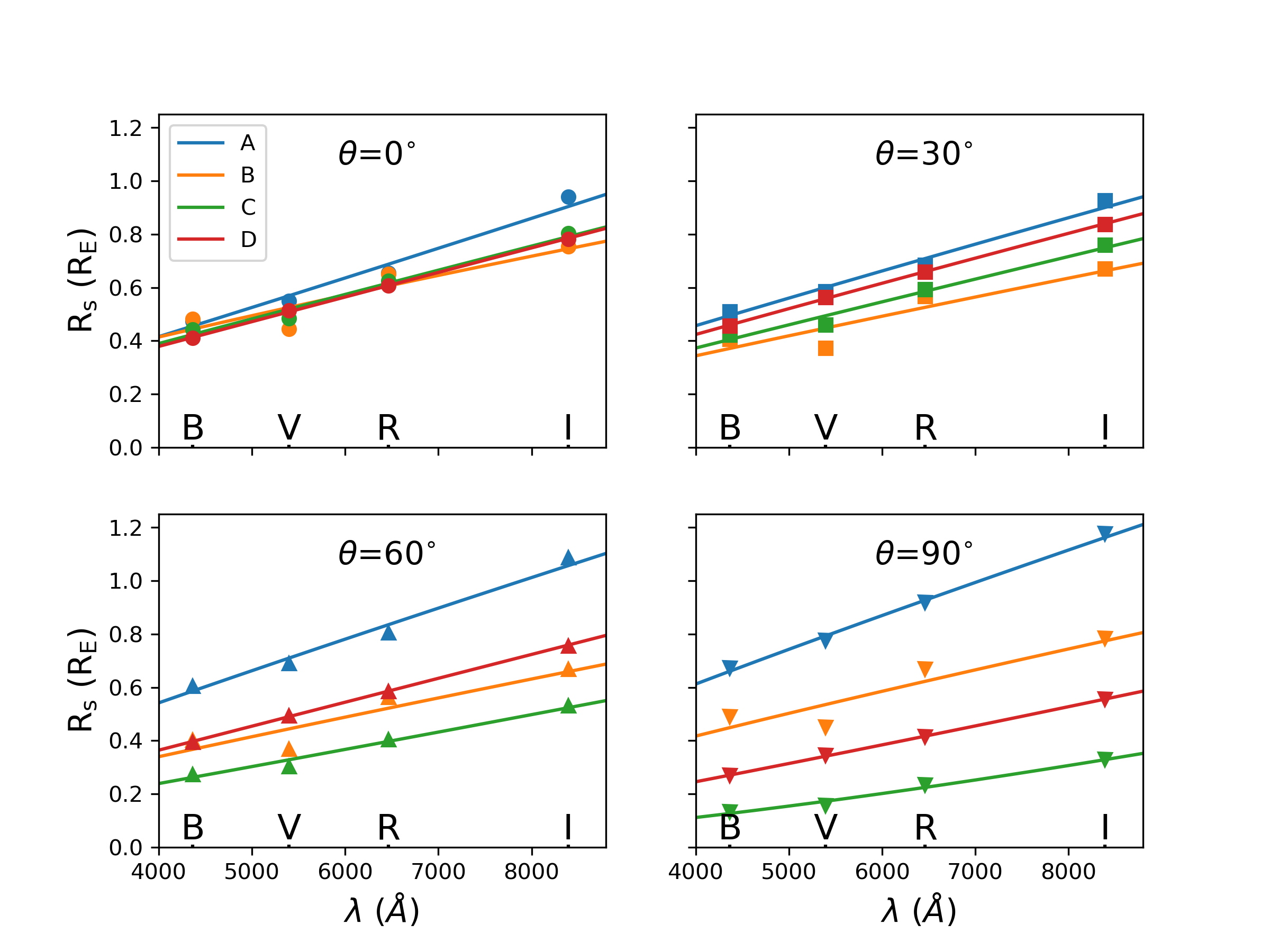}
\caption{Source radius vs. wavelength for four source trajectory angles and the four quasar images. The results for images A, B, C, and D are in blue, orange, green, and red, respectively (see main text).} 
\label{fig:fig11}
\end{figure*}  

In our $g$-band data analysis, central values of $\sigma_{\rm obs}$ are considered (see Table~\ref{tab:sigm}). However, instead of using only these central values, more strictly, small uncertainties of a few percent should be taken into account. Thus, for each value of $\theta$, we also obtained a strip around each $R_{\rm s}$-$L$ curve by searching for pairs ($R_{\rm s}$, $L$) producing $\sigma_{\rm sim}$ values within the measured interval for $\sigma_{\rm obs}$. The four strips (one per quasar image) are much thinner than the separation between them, so they do not modify the estimation of $R_{\rm s}$ by using the central values of the observed dispersions. 

A caveat to our approach is that pairs ($R_{\rm s}$, $L$) coming from $\sigma_{\rm sim}$ values outside the measured intervals for $\sigma_{\rm obs}$ are neglected. Although we expect a small dispersion of standard deviations for a representative set of large simulated trajectories, the dispersion is higher for trajectories of a few Einstein radii. Therefore, values of $\sigma_{\rm sim}$ other than the central values of $\sigma_{\rm obs}$ could give rise to a relatively wide probability band for each image, with probability peaks at the central $R_{\rm s}$-$L$ curves. Here we assume that these probability peaks for each direction in Figure~\ref{fig:fig10} produce reliable uncertainties in the estimation of $R_{\rm s}$, warning that they might be slightly underestimated.

We also note that $L/R_{\rm E} \propto v_{\rm t}/\langle M \rangle^{1/2}$ for a given observation period. Hence, while $L/R_{\rm E}$ depends on $\langle M \rangle^{1/2}$ and the mean mass is not expected to vary much from 0.3 $\rm{M_{\odot}}$, the actual transverse speed could be outside the interval considered here, so the measurement of the source size would be biased. However, although we cannot rule out $v_{\rm t}$ values close to 200 or 700 km s$^{-1}$, their probability is appreciably lower than that of the speed distribution peak. Relative probabilities of $\theta$ = 0 and 90\degr\ are also clearly lower than those of 30$-$60\degr. A detailed interpretation of the observed microlensing signals will be presented in a separate study.

With our $BVRI$ bands long-term optical monitoring data covering the optical spectrum almost uniformly and spanning the same well-sampled period 2004$-$2019, we can also constrain the wavelength dependence of the size of the accretion disc $R_{\rm s} \propto \lambda^{\alpha}$. We followed the procedure detailed above for these $BVRI$ bands. For each optical band, 16 values of $R_{\rm s}$ were derived from the intersections of all $R_{\rm s}$-$L$ curves (corresponding to the four source trajectory angles for each quasar image; see Figure~\ref{fig:fig10} for details) and a horizontal line at the typical trajectory length of 2 $R_{\rm E}$ ($BVRI$ data span 15.5 yr rather than 11.5 yr). These results are included in Figure~\ref{fig:fig11}, where each colour is associated with an image (A in blue, B in orange, C in green, and D in red) and each panel shows results for a trajectory angle. 

Considering all images, we obtain quasi-linear $R_{\rm s}$-$\lambda$ relations with $\alpha$ = 0.92 $\pm$ 0.02 for $\theta$ = 30\degr\ and $\alpha$ = 0.96 $\pm$ 0.07 for $\theta$ = 60\degr. The overall slope is $\alpha$ = 0.97 $\pm$ 0.15 (for all angles and images). A strong point of this analysis is the stability of the slopes against changes in the value adopted for the trajectory length (depending on $v_{\rm t}$). Although changes in the height of the horizontal line describing the adopted value for $L$ (deviations from the typical length; see Figure~\ref{fig:fig10}) produce a systematic increase/decrease of all values of $R_{\rm s}$, the slopes are basically unchanged. 

\section{Conclusions}
\label{sec:end}

The GLENDAMA project has been monitoring a number of 10 gravitationally lensed quasars for the last 20 years with the LT at the Roque de los Muchachos Observatory \citep[La Palma, Spain;][]{Gilm18}. We have presented here a new photometric method to extract magnitudes of the four optical images of the quadruply lensed quasar \object{QSO 2237+0305}, the Einstein Cross, which is a target of the GLENDAMA sample. The new photometric scheme has been applied to LT observations and other monitoring campaigns at the Maidanak Astronomical Observatory (Uzbekistan) and the Cerro Tololo Inter-American Observatory (Chile). All in all, the baseline monitoring time of the Einstein Cross extends 30 years and the associated light curves describe the variability in the $BgVrRI$ optical bands. This represents an enormous effort in treating and combining data from a wide variety of instruments and observing conditions.

As a result, DLCs are provided within the new global set of data (Figure~\ref{fig:fig7}), where the extrinsic variability due to microlensing effects at the position of the quasar images, can be analysed with high accuracy. Note that microlensing signals exhibit sharp variations that depend on wavelength, making them very promising features for resolving the controversy over the structure of the quasar accretion disc \citep[e.g.,][]{Eige08b,Mun16,Goi20}. Although long-term optical monitorings are usually conducted in one band-pass filter, and thus they cannot be used to constrain the accretion disc structure, the source radius-wavelength relationship $R_{\rm s} \propto \lambda^{\alpha}$ can be probed with our long-term follow-up of the variability of \object{QSO 2237+0305} from the blue to the near-infrared part of its spectrum. Since a detailed analysis of the new DLCs from microlensing simulations was beyond the scope of this work, that will be presented in a forthcoming paper using large high-resolution magnification maps.

However, a preliminary study of the accretion disc structure is performed solely using the root mean square dispersions of the DLCs over well-sampled time periods (Table~\ref{tab:sigm}). These wavelength-dependent standard deviations for each quasar image have been compared to those from simulated DLCs for Gaussian source trajectories across dynamic magnification maps. Our simulated DLCs account for stellar random motions and are sampled as the observed ones. Adopting a mean microlens mass $\langle M \rangle$ = 0.3 $\rm{M_{\odot}}$ and reasonable constraints on the effective velocity of the source in the lens plane \citep[400 $\pm$ 60 km s$^{-1}$ with a position angle within the interval 30$-$60\degr;][]{Poin10a}, we have measured the half-light radius $R_s$ of the $g$-band source (Figure~\ref{fig:fig10}) and derived a power-law index of $\alpha \sim$ 1 (Figure~\ref{fig:fig11}) rather than the value predicted by the geometrically-thin, optically-thick standard disc theory \citep[$\alpha$ = 4/3;][]{Shak73}. Microlensing analyses of other lensed quasars also found slopes shallower than the standard one \citep[e.g.,][]{Blac11}, whereas \citet{Corn20} generally found slopes steeper than the standard model from quasar variability studies (temperature profiles shallower than the standard one). 

For the $g$-band source, $R_s$ is found to be in the range of 7 to 12 lt-day, in very good agreement with the measurements of \citet{Mun16} at $\sim$4700 \AA\ for Gaussian, hybrid and standard-disc sources (scaled to $\langle M \rangle$ = 0.3 $\rm{M_{\odot}}$), but somewhat larger than the upper limit of $\sim$6 lt-day on the $g$-band half-light radius from the analysis of the high-magnification event in the image C in 1999 (see Sect.~\ref{sec:phot}), using a Gaussian source model without any prior on the velocity of the lens galaxy \citep{Angu08}. In addition, our results indicate a source size that grows with wavelength with a power-law index (slope) of $\alpha$ = 0.94 $\pm$ 0.05 (averaging the central values and errors for position angles of 30 and 60\degr). This slope is consistent with those reported by \citet{Eige08b} and \citet{Mun16}, but shallower than the standard disc model at $\sim$8$\sigma$ significance. Even assuming a possible underestimation of the error in $\alpha$ by a factor of two or three, there are serious difficulties for a standard accretion disc as the sole source of UV-optical continuum radiation in \object{QSO 2237+0305}. \citet{Gask08} suggested a model with a much more inconsistent slope ($\alpha$ = 1.75) than the standard value, while $\alpha$ = 8/7 if the disc is powered by the spin of the central black hole \citep{Agol00}. Although a non-standard accretion disc model can reproduce the observations, optical passband fluxes come from the central accretion disc and the broad emission-line region \citep[e.g.,][]{Kor01}, so the measured relationship between source size and wavelength could be considerably flattened if the contribution of the extended region is substantial \citep[e.g.,][]{Fian23}. 

This result from a long-term multi-band photometric monitoring opens a window for crucial tests of the quasar structure and demonstrates that microlensing is able to reveal the possible deviations from a standard accretion disc scenario. Further analysis of other lensed quasars of the GLENDAMA sample, as well as implementing larger, high-resolution magnification map cubes for the full light curves would definitely provide ultimate constraints to compare with results obtained in this and other works.

\section*{Data availability}
\label{sec:data}

Tables C.1-C.6 are only available in electronic form at the CDS via anonymous ftp to cdsarc.u-strasbg.fr (130.79.128.5) or via http://cdsweb.u-strasbg.fr/cgi-bin/qcat?J/A+A/vol/id
     
\begin{acknowledgements}
We thank Christopher Kochanek (CK) for his comments and suggestions, which have helped us to significantly improve an initial version of the manuscript. The Liverpool Telescope (LT) is operated on the island of La Palma by Liverpool John Moores University in the Spanish Observatorio del Roque de los Muchachos of the Instituto de Astrof\'isica de Canarias with financial support from the UK Science and Technology Facilities Council. We thank the staff of the LT for a kind interaction. The 1.5 m Maidanak Telescope is a facility of the Ulugh Beg Astronomical Institute of the Uzbekistan Academy of Sciences, which is operated on the Maidanak Astronomical Observatory (MAO) by a consortium of various international institutions. We thank Boris Artamonov, Otabek Burkhonov, Talat Akhunov, Ildar Asfandiyarov, Vasily Bruevich, and Shuhrat Ehgamberdiev, who organized and carried out the observations of the Einstein Cross at the MAO. The 1.3 m SMARTS Telescope is operated on the Cerro Tololo Inter-American Observatory by the SMARTS Consortium, and the Einstein Cross monitoring program was led by CK. VNS acknowledges the Universidad de Cantabria (UC) and the Spanish Agencia Estatal de Investigacion (AEI) for financial support for a long stay at the UC in the period 2022–2024. AEG acknowledges support from project ANID Fondecyt Postdoctorado with grant number 3230554. This research was supported by the grant PID2020-118687GB-C31, financed by the Spanish Ministerio de Ciencia e Innovación through MCIN/AEI/10.13039/501100011033. This research has also been supported by UC funds and the grant PID2020-118990GB-I00 funded by MCIN/AEI/10.13039/501100011033. 
\end{acknowledgements}

\clearpage
\appendix
\onecolumn
\section{Zoomed-in multi-band light curves}
\FloatBarrier
\label{sec:zoom} 
\begin{figure*}[h!]
\centering
\includegraphics[width=9cm]{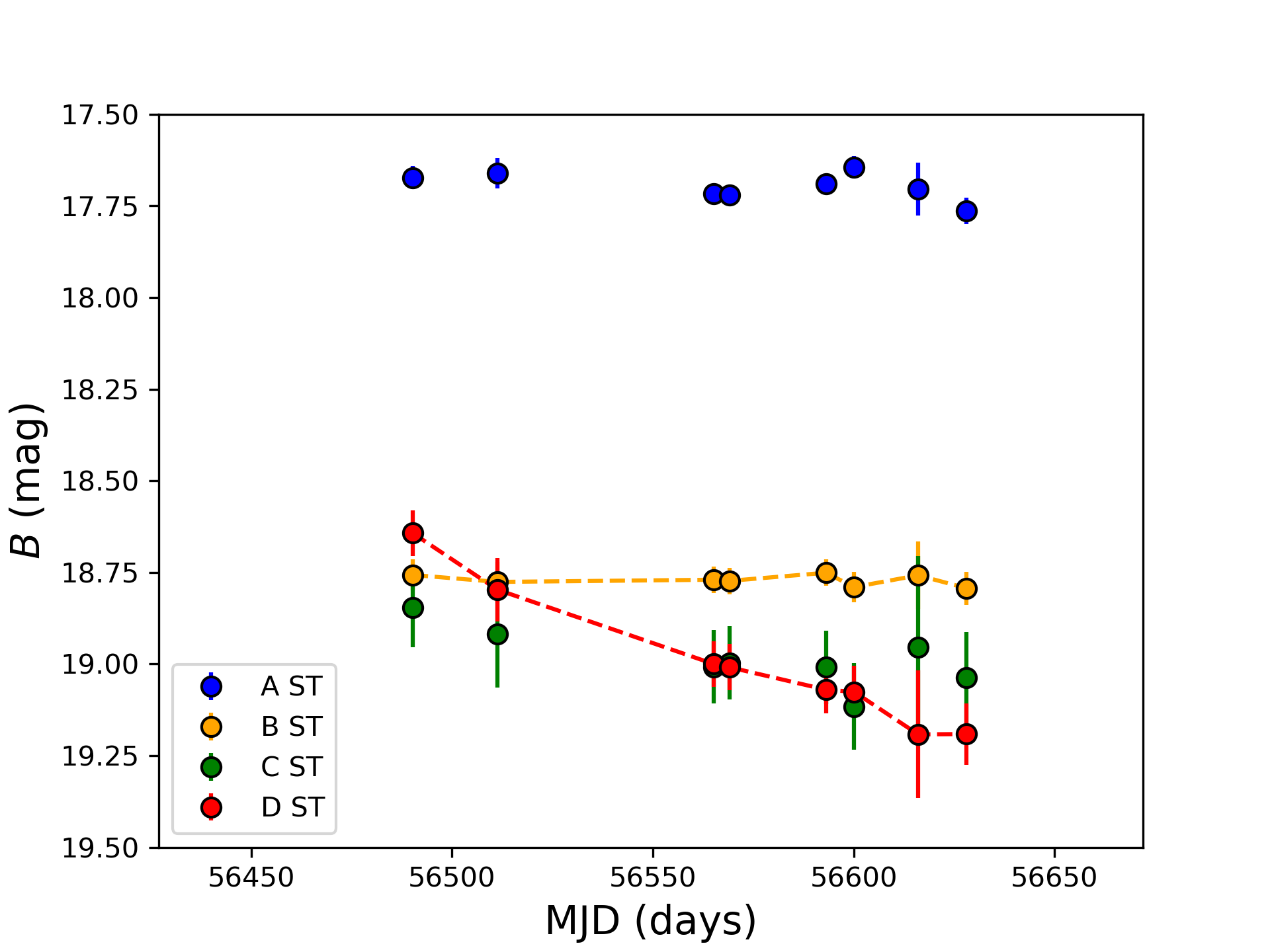}
\includegraphics[width=9cm]{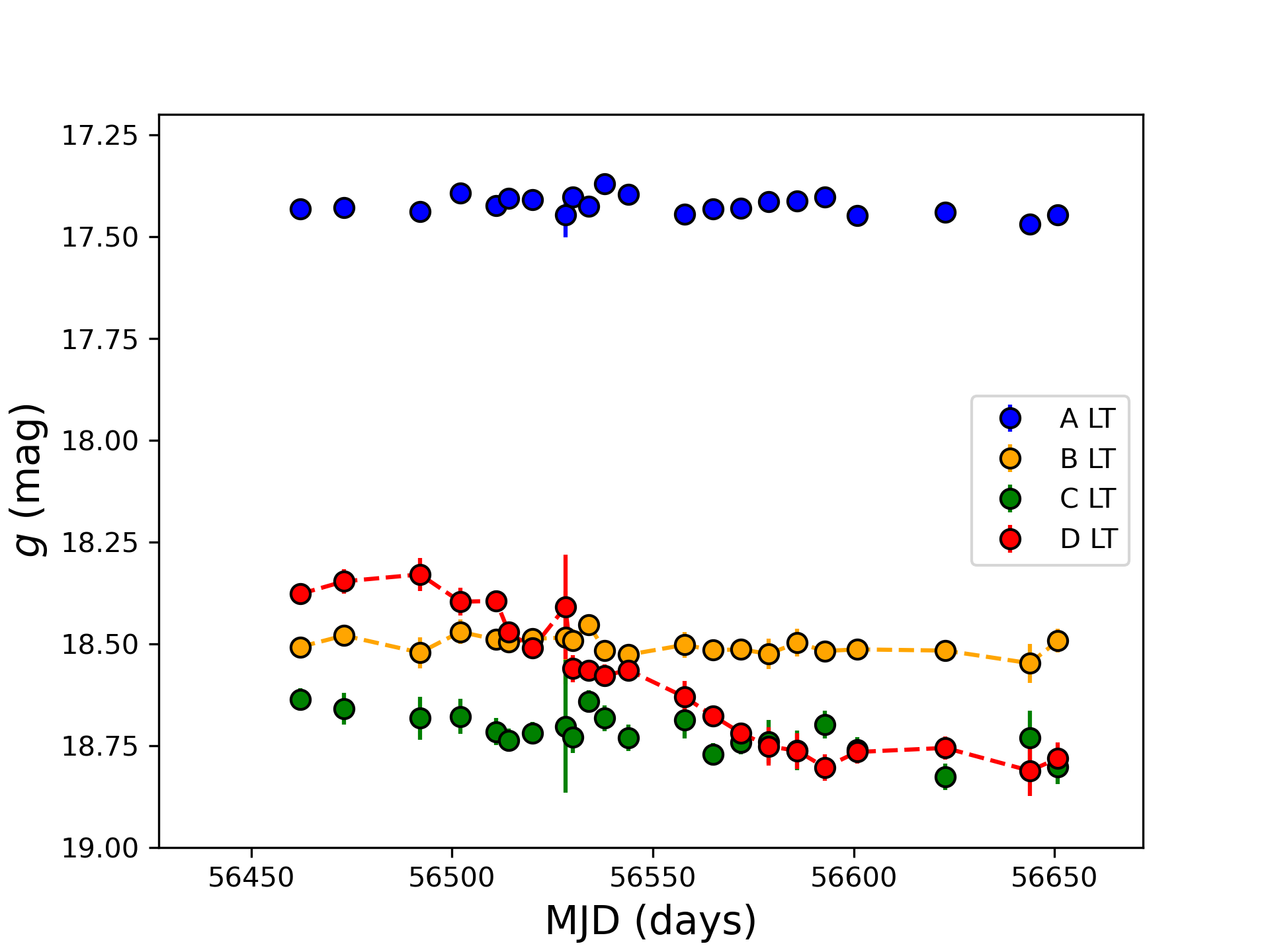}
\includegraphics[width=9cm]{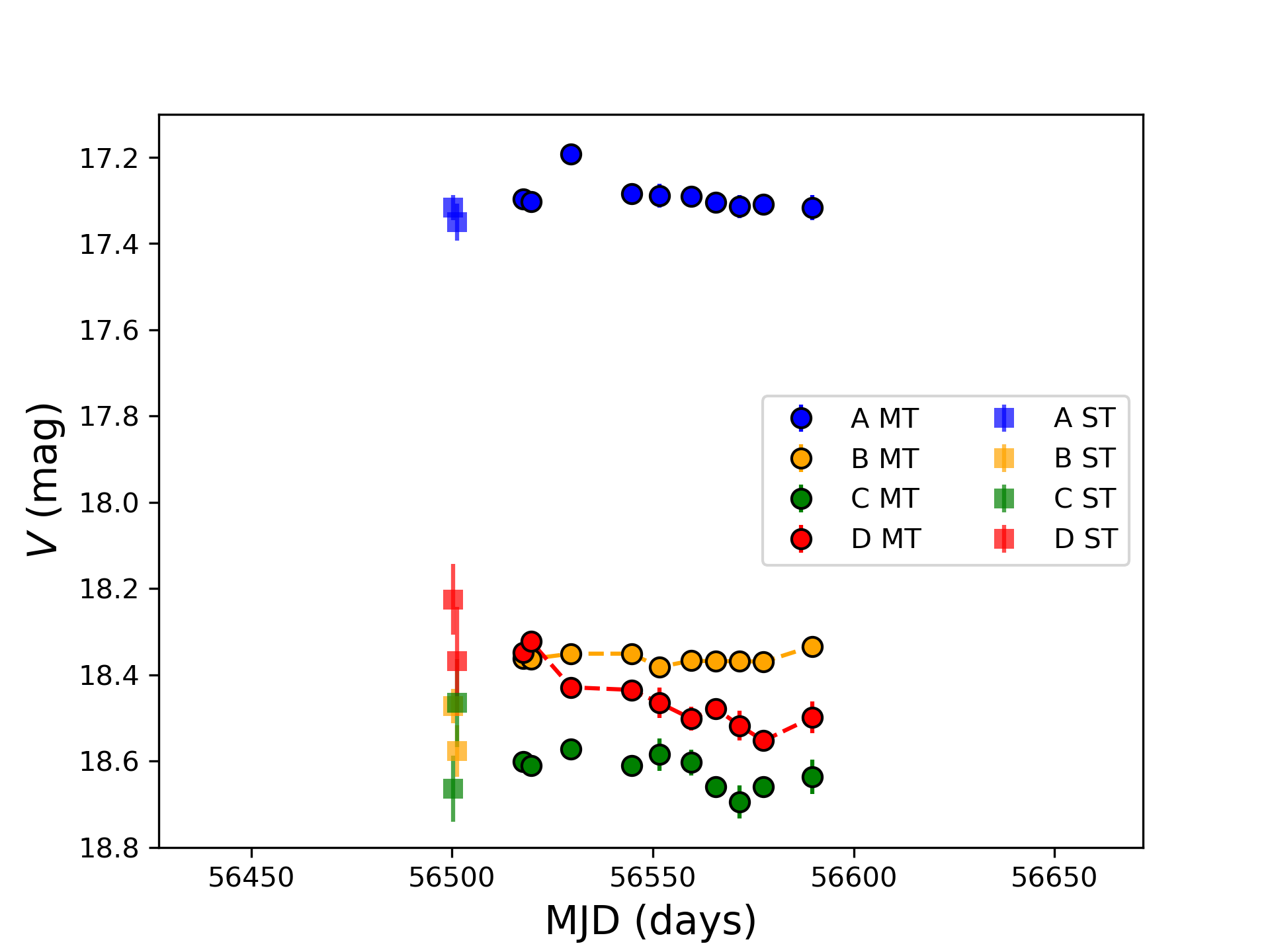}
\includegraphics[width=9cm]{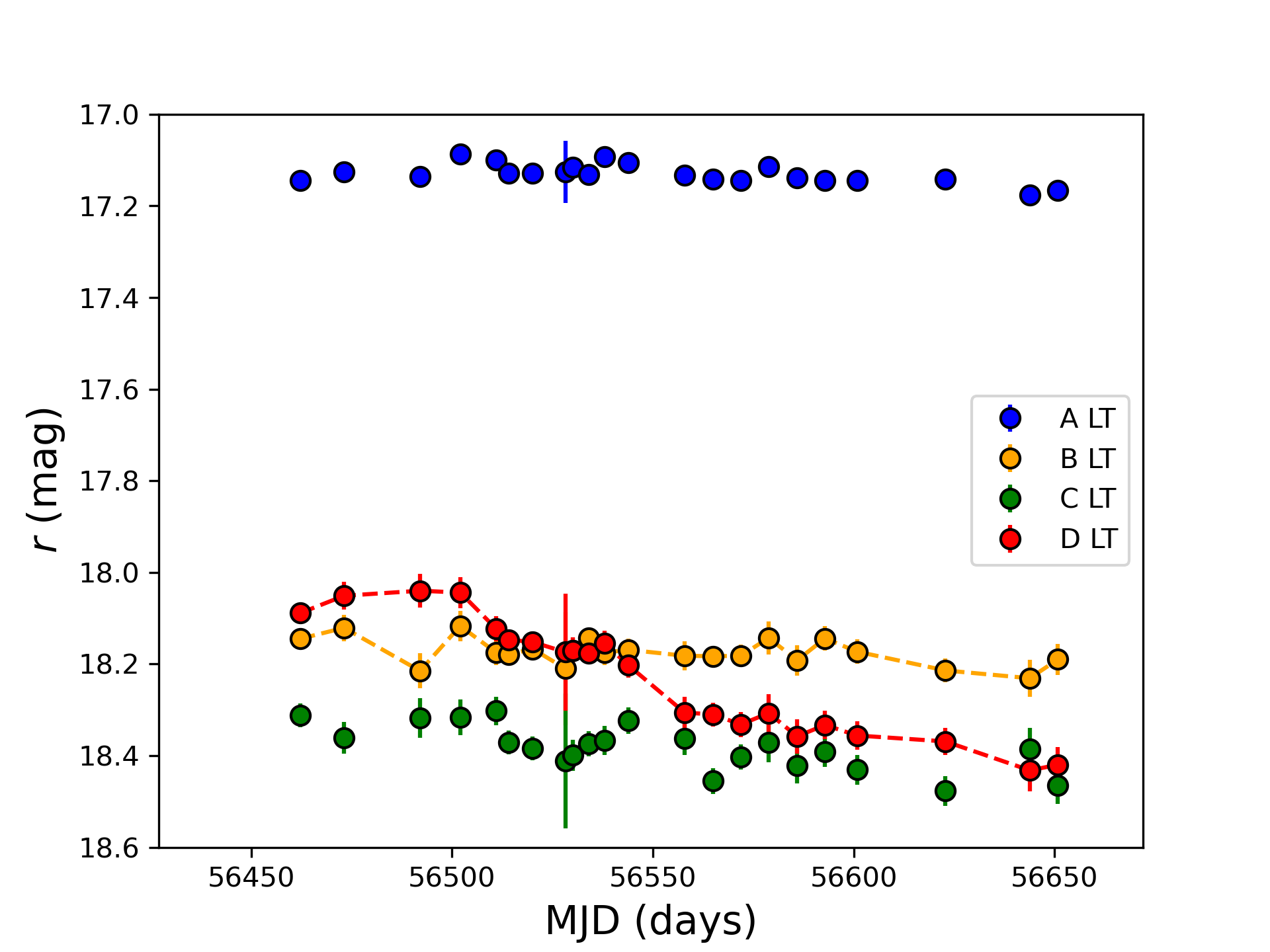}
\includegraphics[width=9cm]{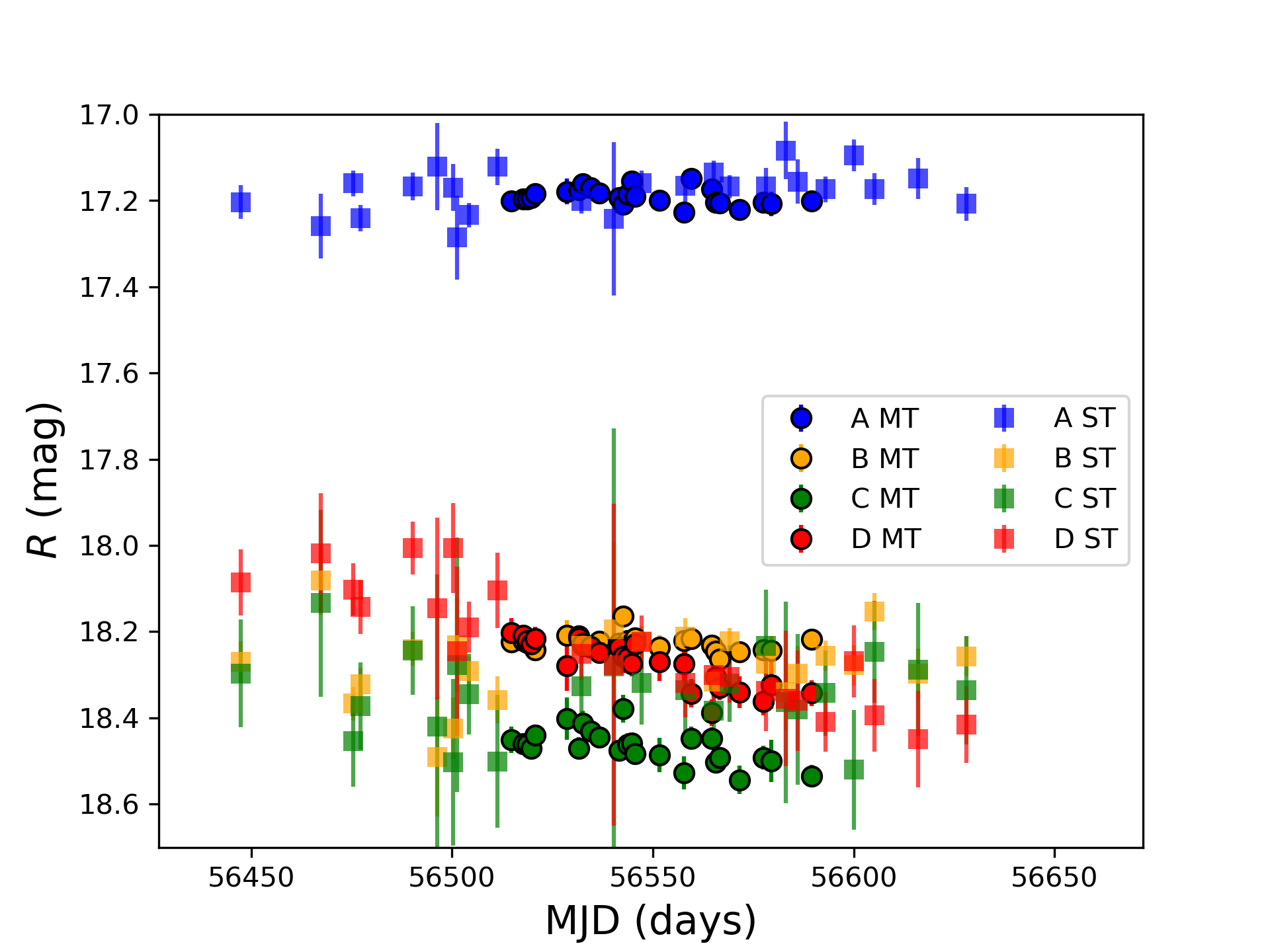}
\includegraphics[width=9cm]{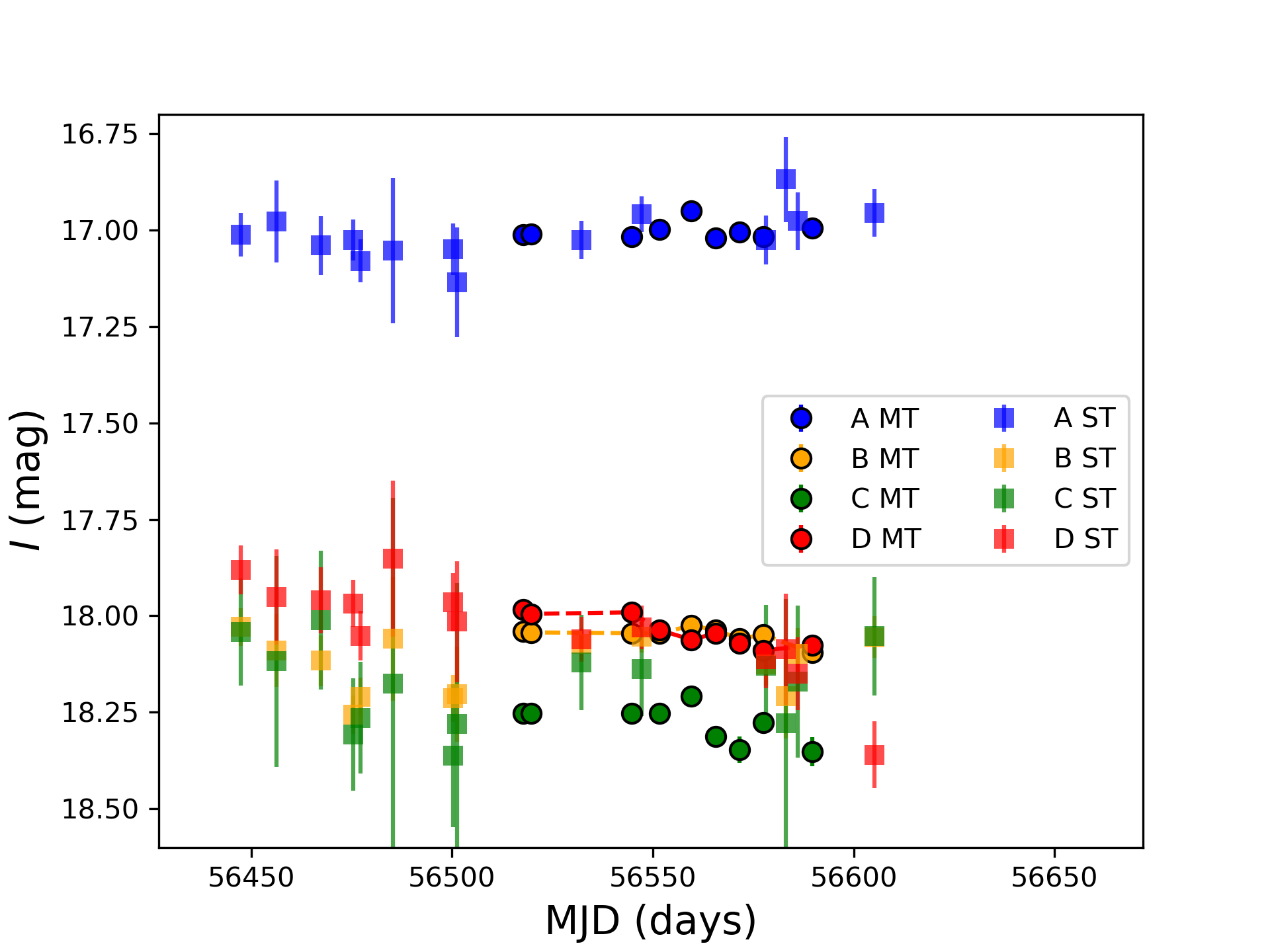}
\caption{Multi-band light curves in 2013. Regarding the variability of images B and D, we show dashed lines as a guide to the eye.}
\label{fig:lcs2013}
\end{figure*} 
\newpage
\begin{figure*}[!t]
\centering
\includegraphics[width=9cm]{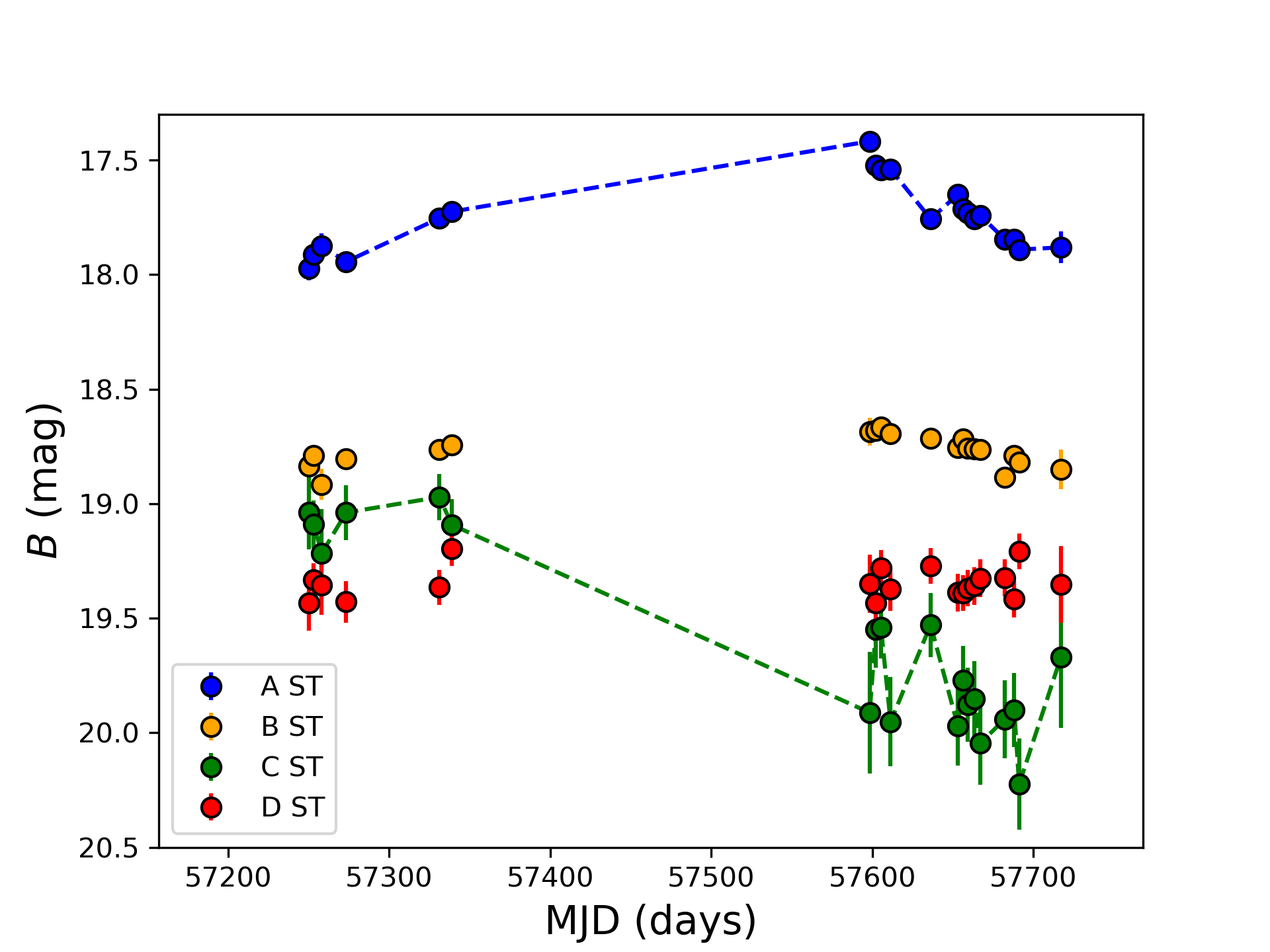}
\includegraphics[width=9cm]{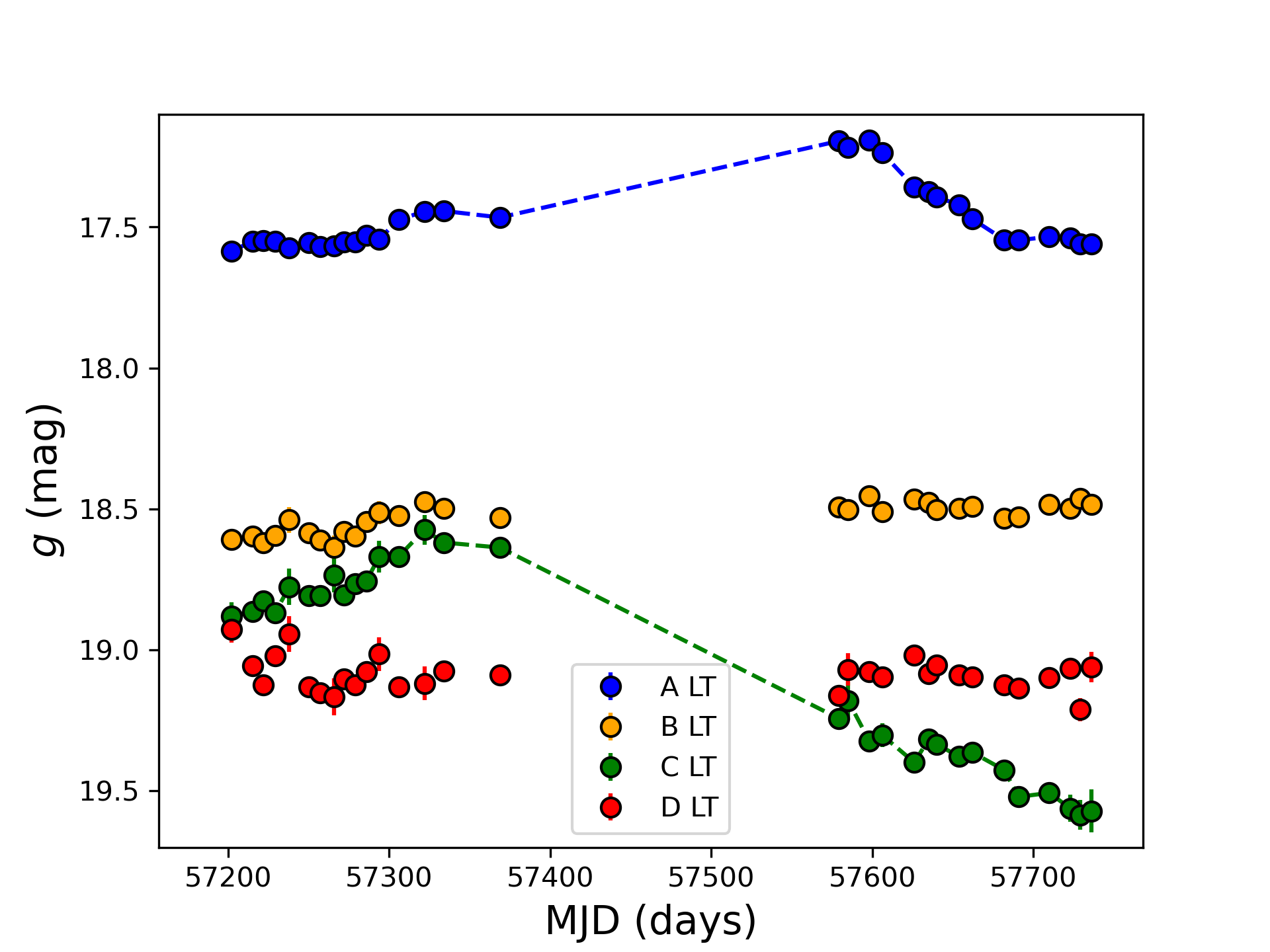}
\includegraphics[width=9cm]{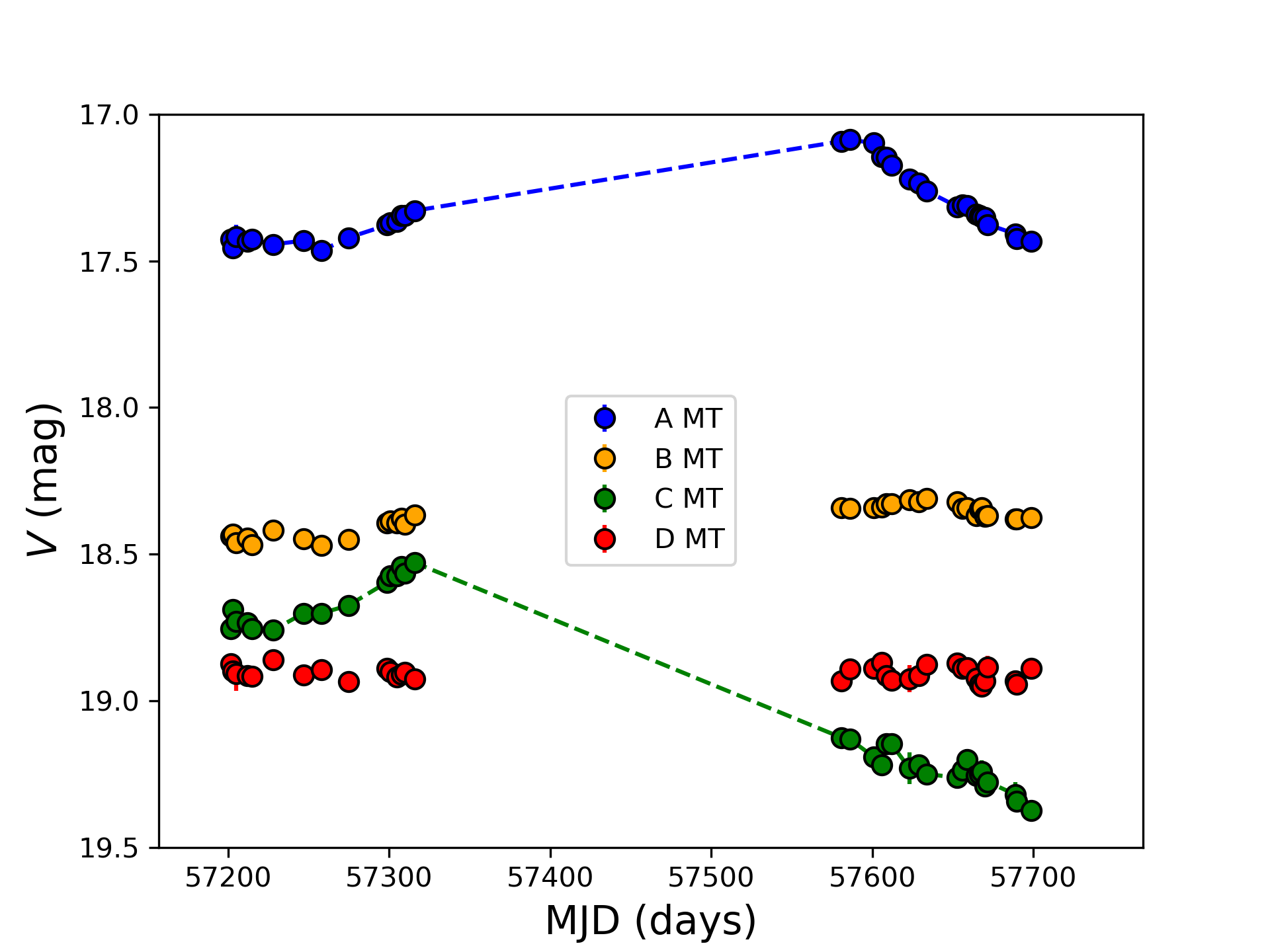}
\includegraphics[width=9cm]{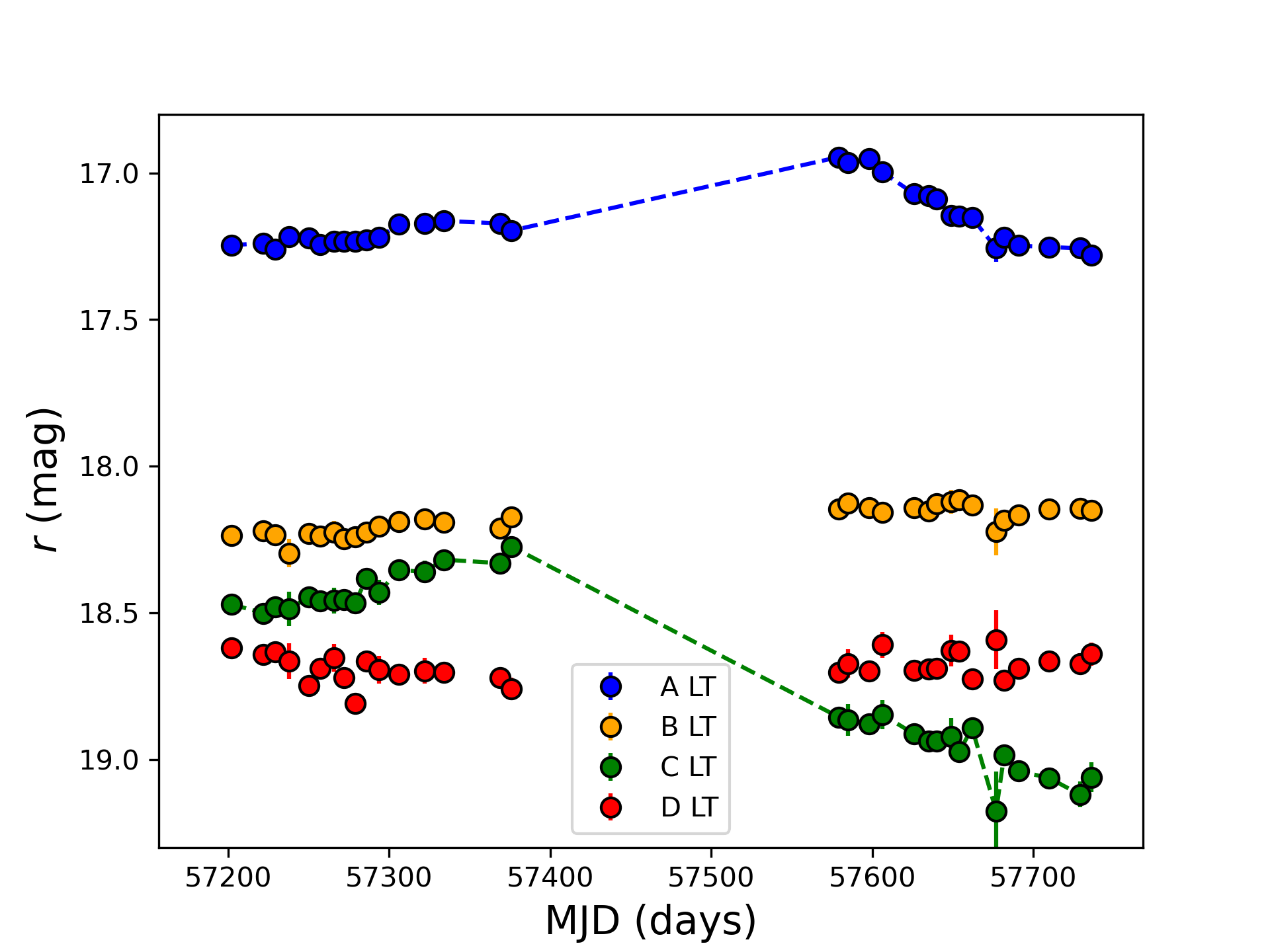}
\includegraphics[width=9cm]{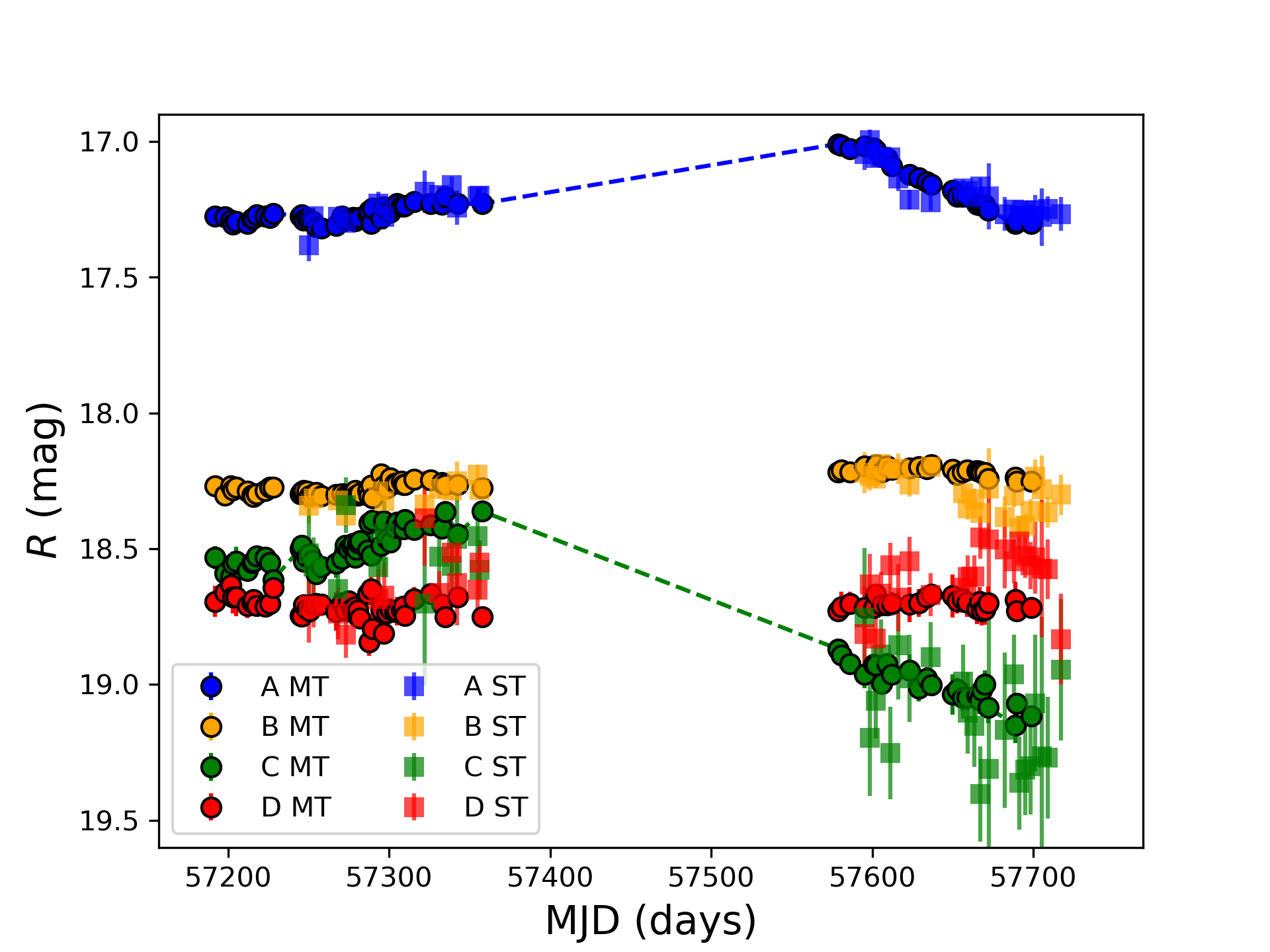}
\includegraphics[width=9cm]{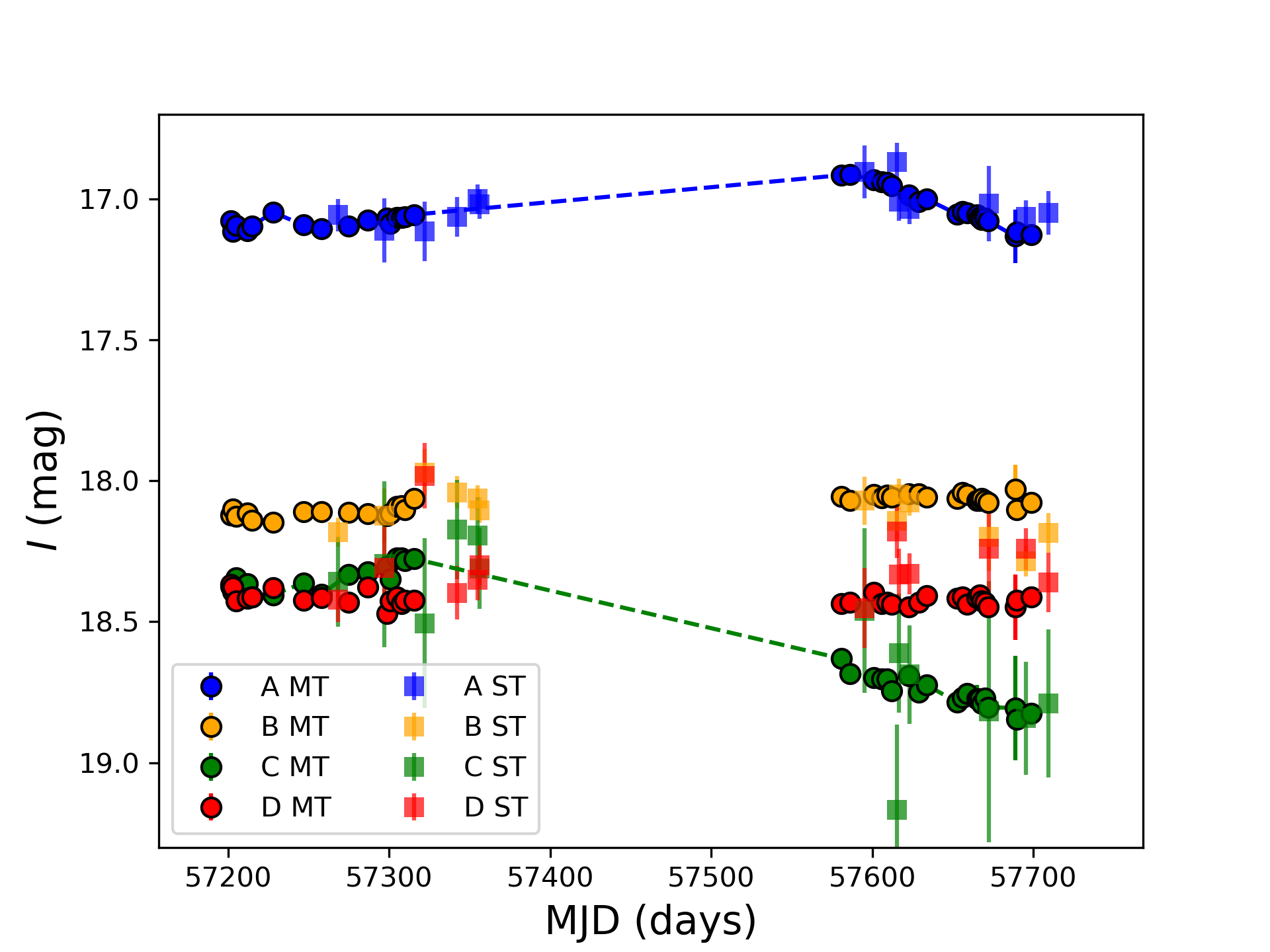}
\caption{Multi-band light curves in 2015$-$2016. We include dashed lines as a guide to the eye.}
\label{fig:lcs2015_16}
\end{figure*} 
\newpage
\begin{figure*}[!t]
\centering
\includegraphics[width=9cm]{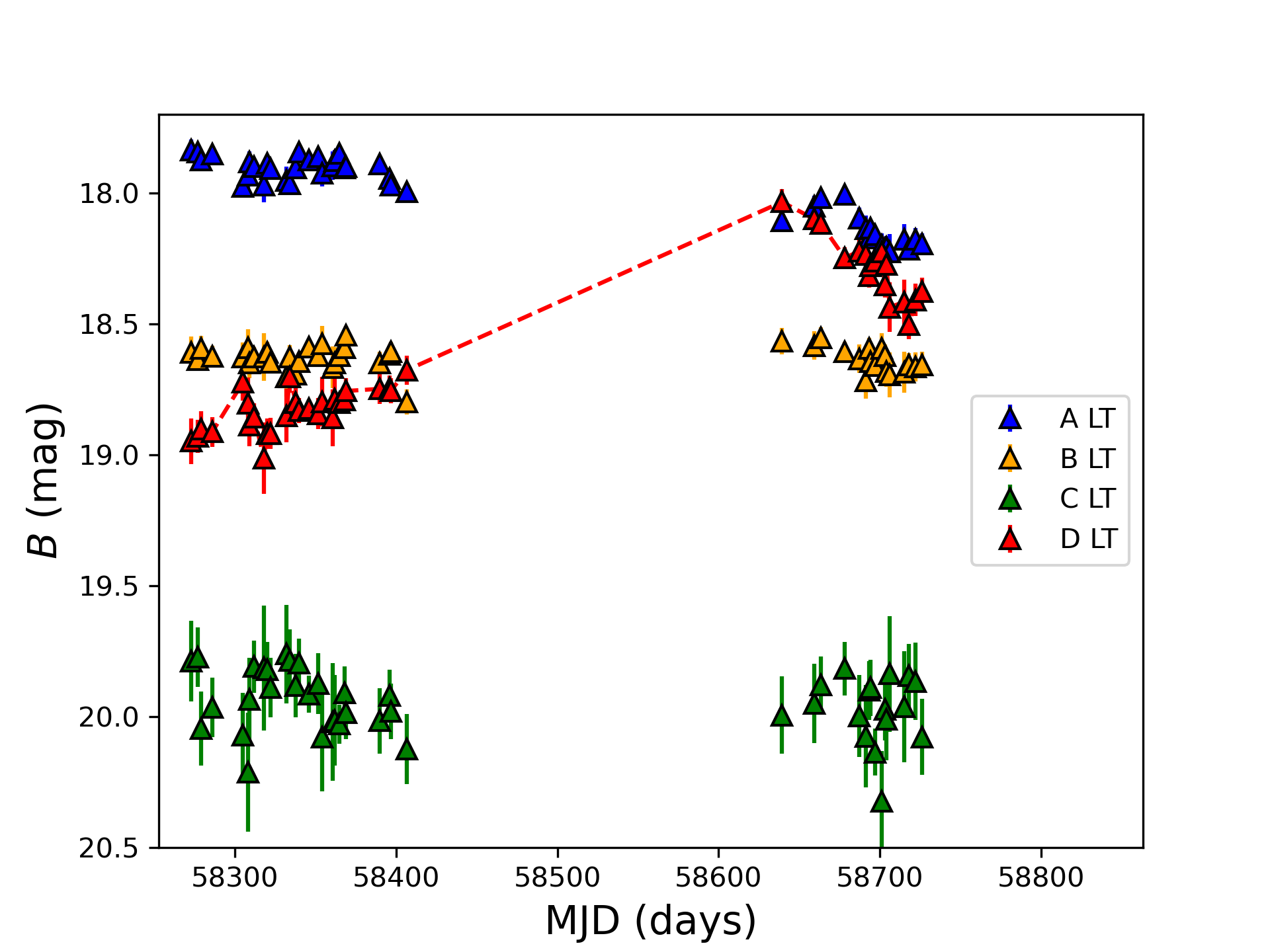}
\includegraphics[width=9cm]{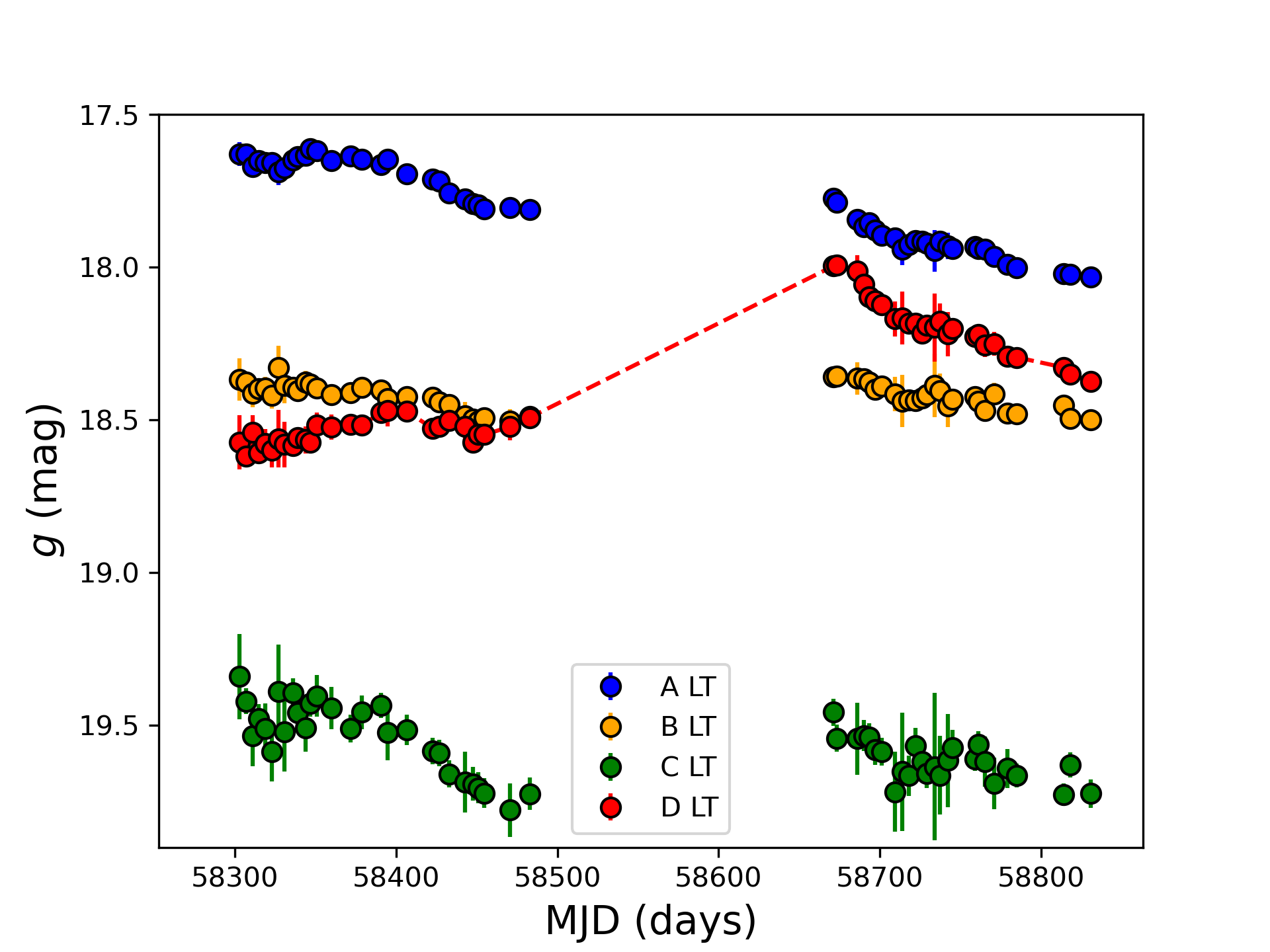}
\includegraphics[width=9cm]{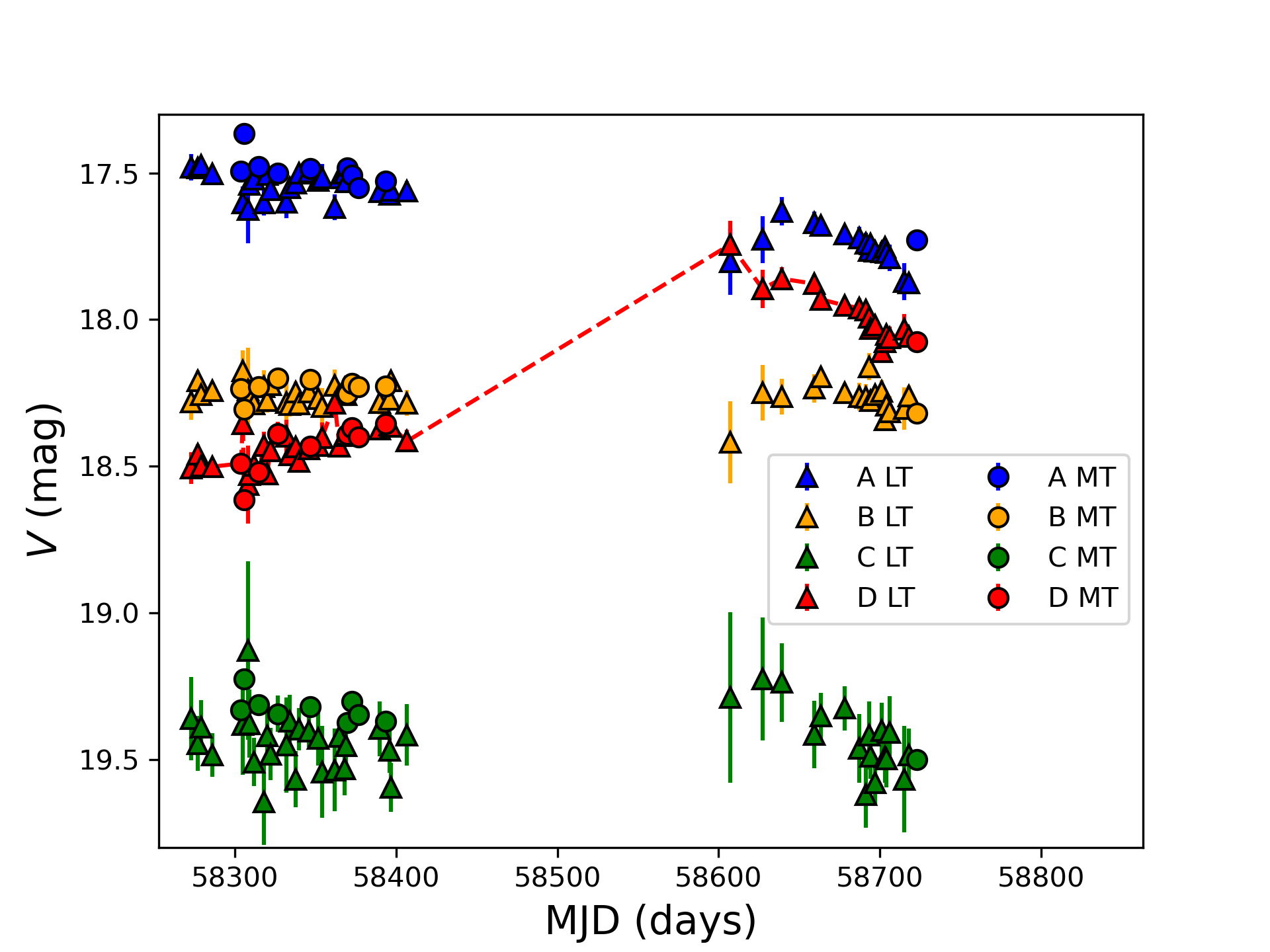}
\includegraphics[width=9cm]{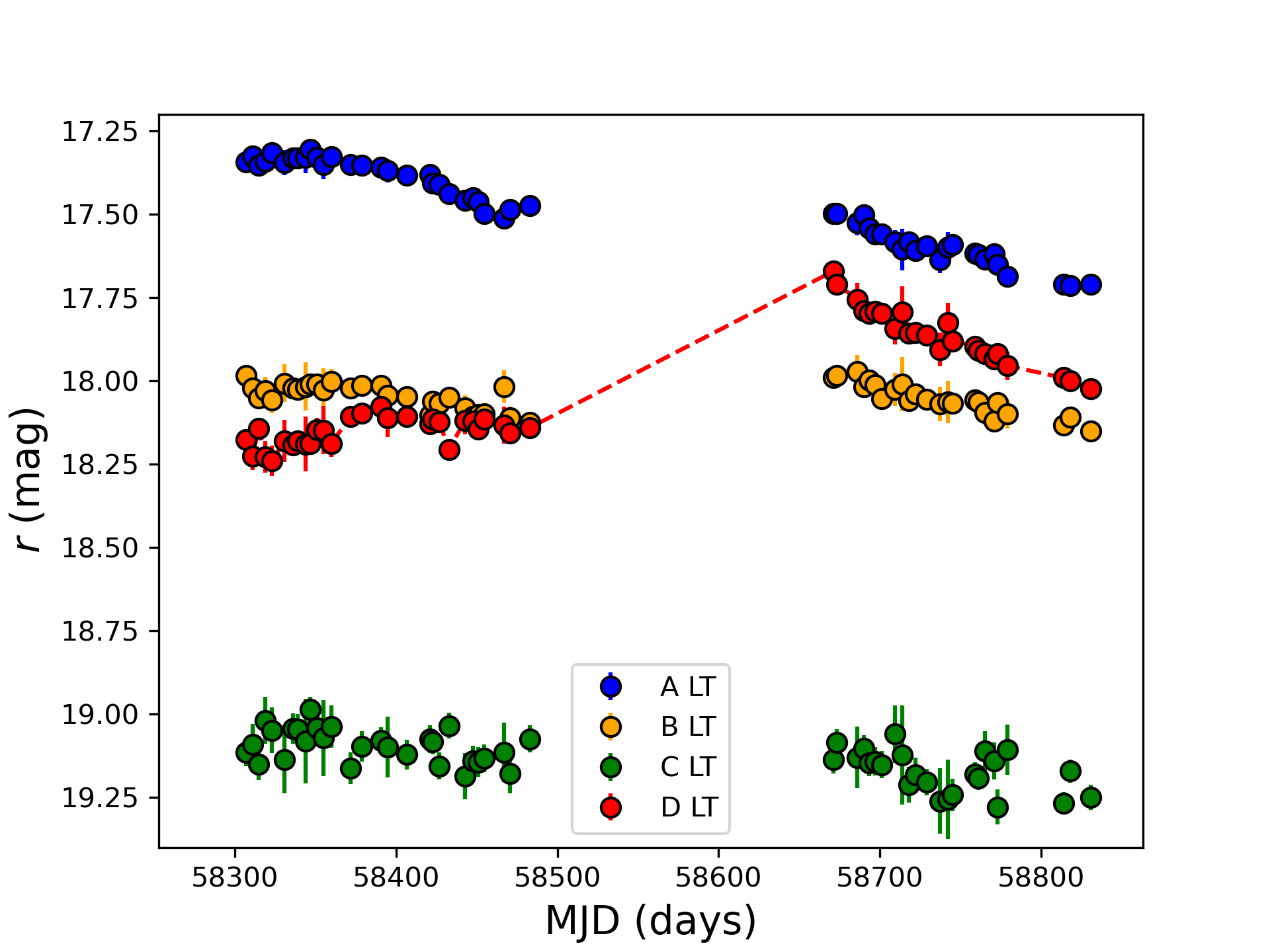}
\includegraphics[width=9cm]{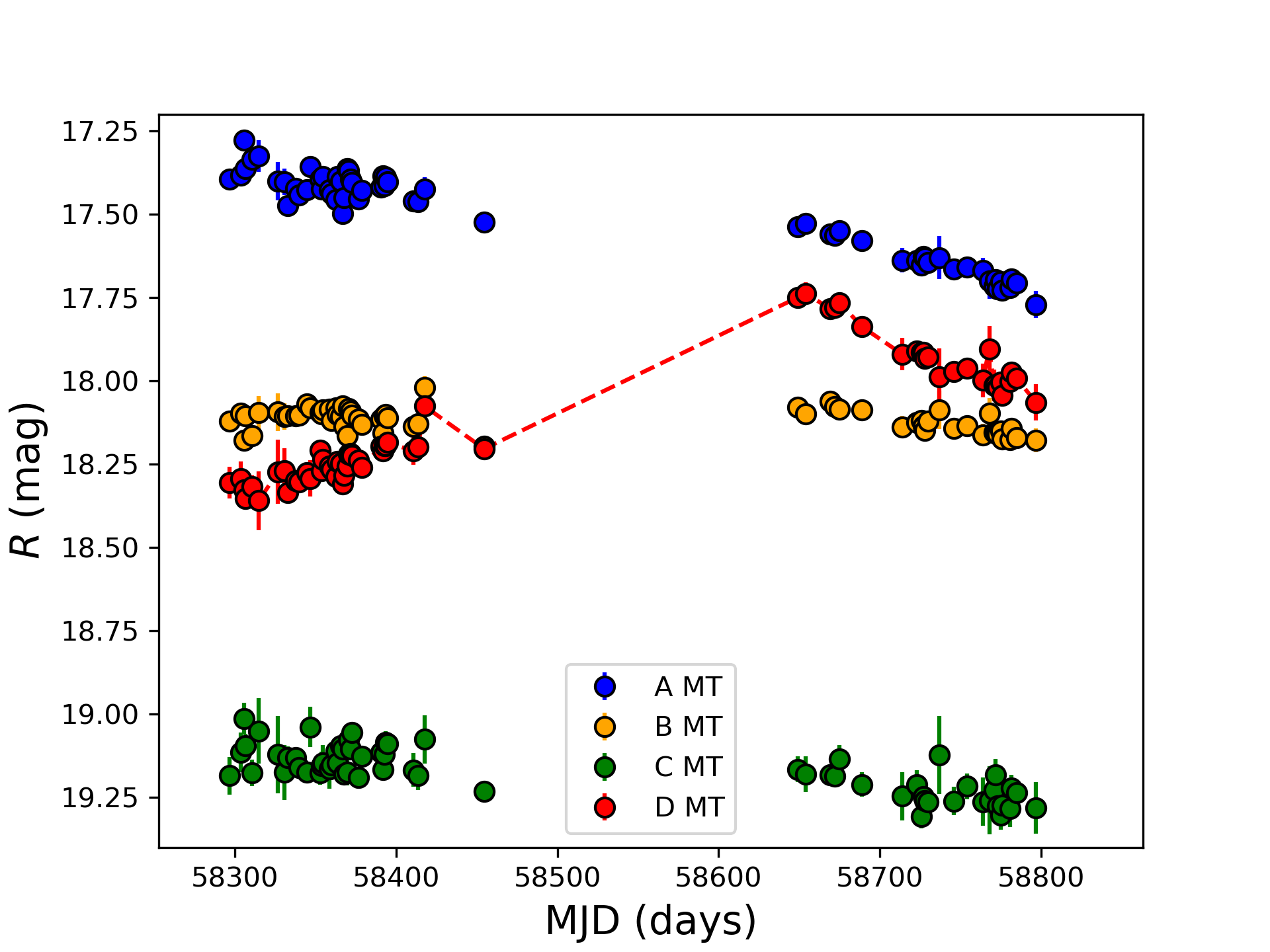}
\includegraphics[width=9cm]{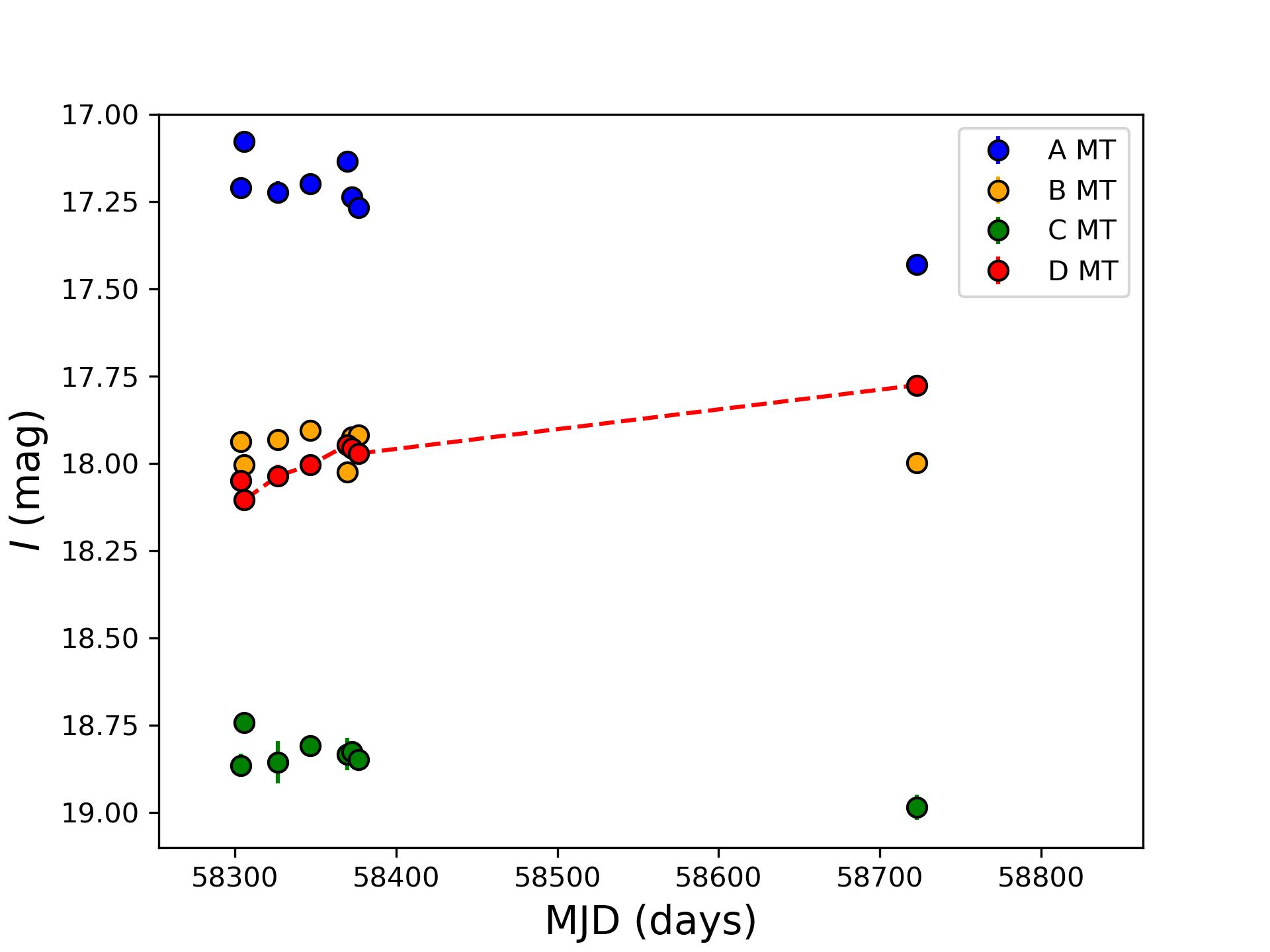}
\caption{Multi-band light curves in 2018$-$2019. We include dashed lines as a guide to the eye.}
\label{fig:lcs2018_19}
\end{figure*} 

\onecolumn
\section{Zoomed-in multi-band DLCs}
\FloatBarrier
\label{sec:zoom2}
\begin{figure*}[h!]
\centering
\includegraphics[width=0.85\textwidth]{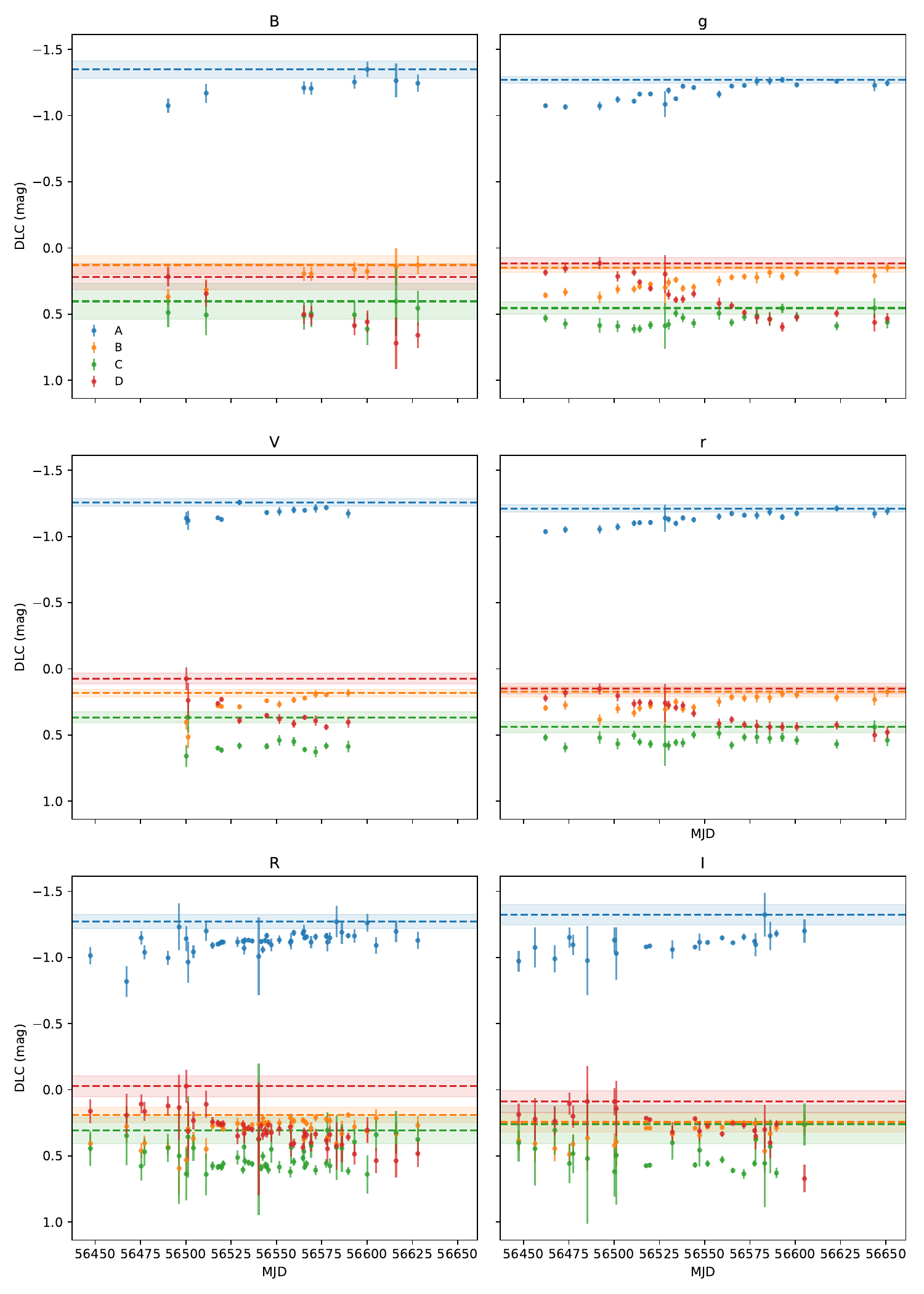}
\caption{Multi-band DLCs in 2013. The horizontal dashed lines and the associated horizontal strips represent the average error bars for the four images in the six bands.}
\label{fig:dlcs2013}
\end{figure*} 
\newpage
\begin{figure*}[!t]
\centering
\includegraphics[width=0.85\textwidth]{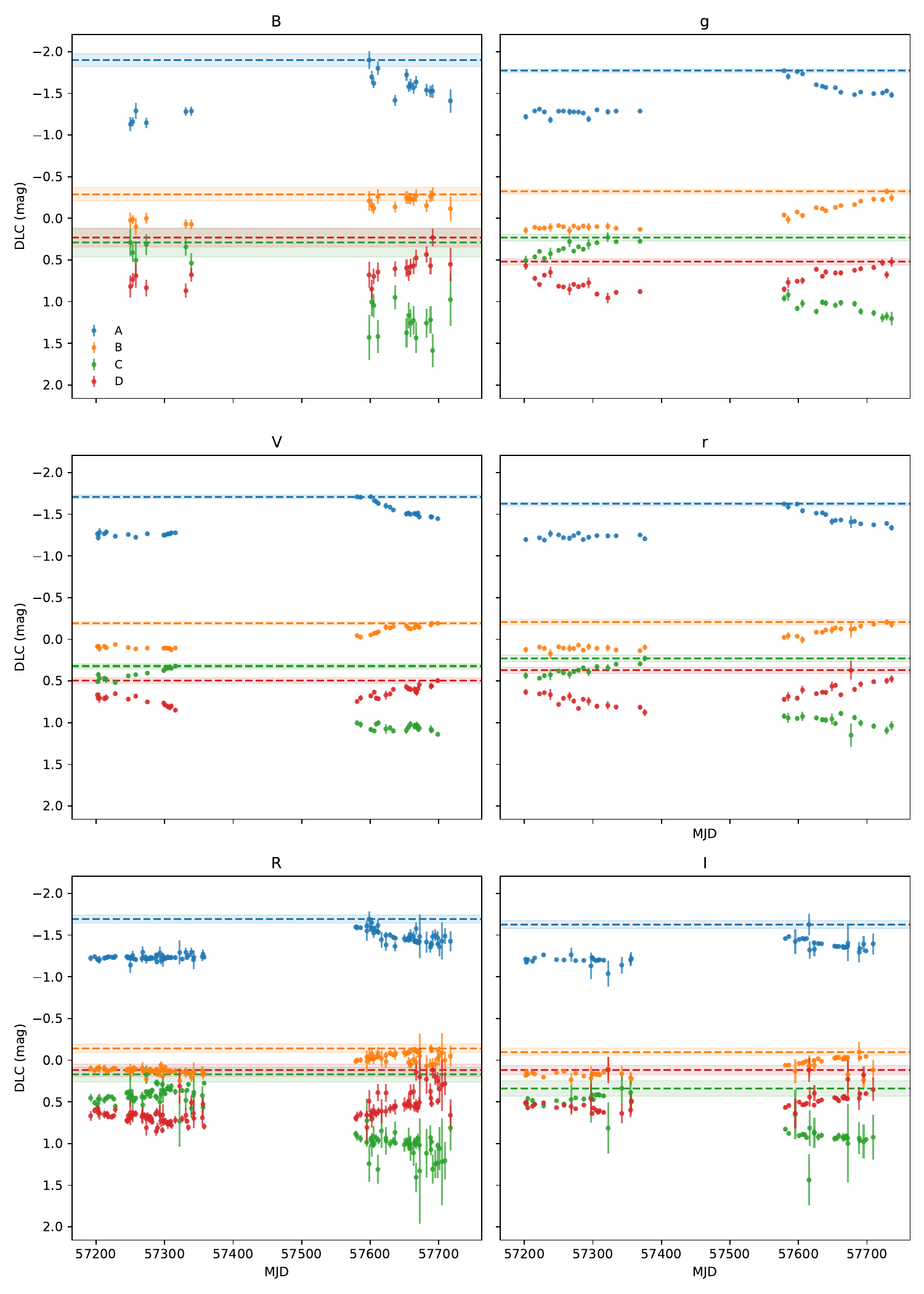}
\caption{Multi-band DLCs in 2015$-$2016. The horizontal dashed lines and the associated horizontal strips represent the average error bars for the four images in the six bands.}
\label{fig:dlcs2013}
\end{figure*} 
\newpage
\begin{figure*}[!t]
\centering
\includegraphics[width=0.85\textwidth]{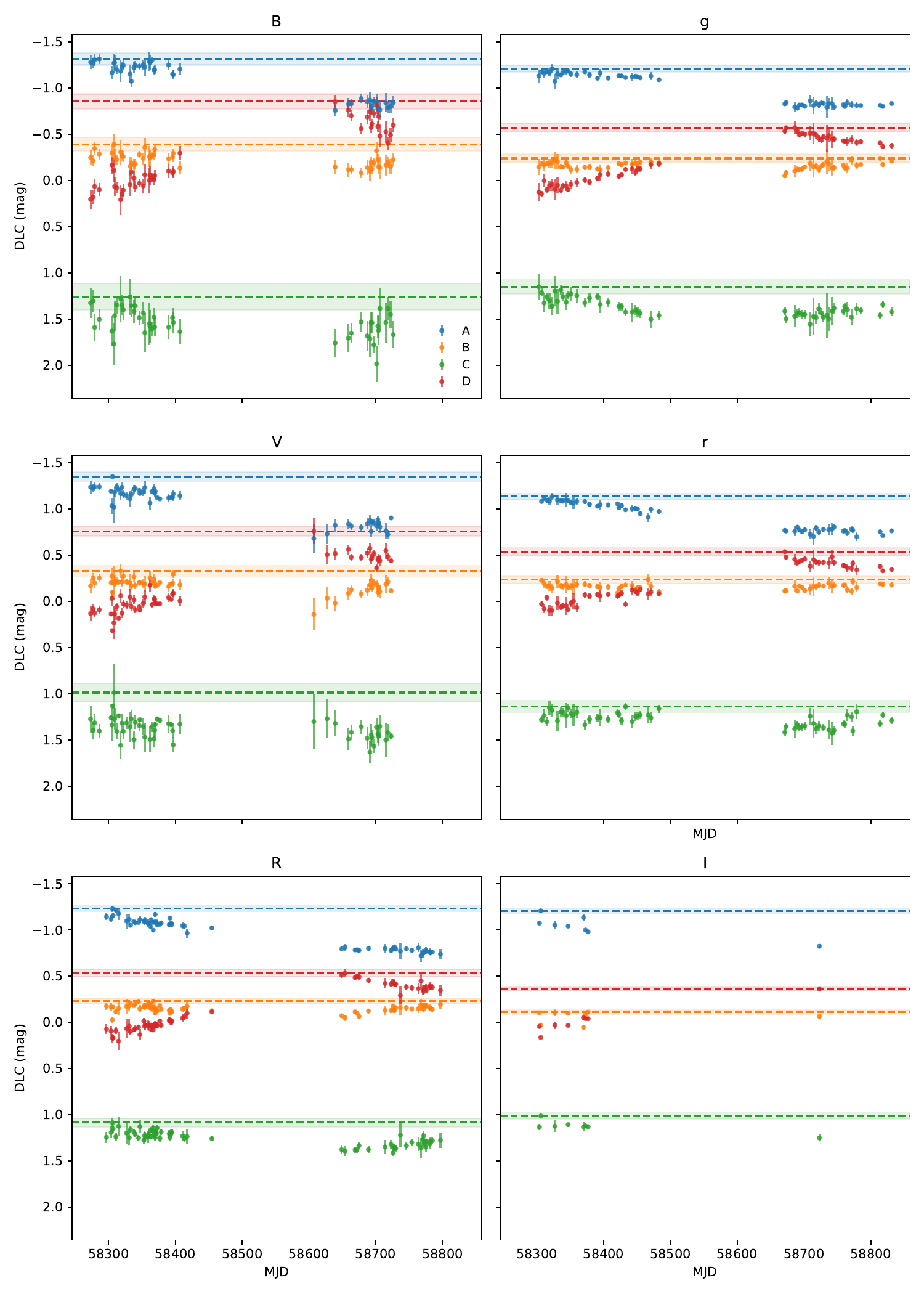}
\caption{Multi-band DLCs in 2018$-$2019. The horizontal dashed lines and the associated horizontal strips represent the average error bars for the four images in the six bands.}
\label{fig:dlcs2013}
\end{figure*} 

\end{document}